\documentclass[a4paper,dvips,12pt]{article}
\usepackage{mcite}
\input{axodraw.sty}
\usepackage{amsmath,amssymb,exscale}   
\usepackage{array,multicol}
\usepackage{afterpage,float,flafter}   
\usepackage{epsfig,rotating,pifont,fancybox}   
\usepackage{cite}
\usepackage{axodraw}
\makeatletter   
\@addtoreset{equation}{section}   
\makeatother   
   
\newcommand{\bea}{\begin{eqnarray}}   
\newcommand{\eea}{\end{eqnarray}}

\newcommand{\Tr}{\mathop{\rm Tr}}

\newcommand\One{\mbox{\rm 1\hspace{-3pt}I}}  

\def\ctu(#1,#2){
\BCirc(#1,#2)4
\put(0,0){
}
\put(0,0){
}
}
\def\ctq(#1,#2){
\BCirc(#1,#2)4
\put(0,0){
}
\put(0,0){
}
}

\def\be{\begin{equation}}
\def\ee{\end{equation}}
\def\a{\alpha}

\def\g{\gamma}

\def\d{\delta}

\def\r{\rho}

\def\l{\lambda}
\def\L{\Lambda}
\def\m{\mu}
\def\n{\nu}
\def\o{\omega}

\def\s{\sigma}

\newcommand{\brr}{\begin{eqnarray}}
\newcommand{\err}{\end{eqnarray}}
\newcommand{\ns}{,\qquad \text{(no sum)}}
\newcommand{\fig}[1]{Fig.~\ref{#1}}

\title{   
\vspace*{-0.8cm}   
\begin{flushright}   
\normalsize{      
IEM-FT-223/02\\
IFT-UAM/CSIC-02-11\\   
\texttt{hep-th/0204223}}\\ 
\end{flushright}    
\vspace{1cm}
\Large{\sc Bulk and brane radiative effects in gauge theories 
on orbifolds~\footnote{Work 
supported in part by CICYT, Spain, under contract AEN98-0816,
and by EU under contracts HPRN-CT-2000-00152 and HPRN-CT-2000-00148.}}
\vspace*{.5cm}
\author{\large
{\sc G.~v.~Gersdorff, N.~Irges  and M.~Quir{\'o}s}\\ \\
\emph{Instituto de Estructura de la Materia (CSIC), Serrano 123,}\\
\emph{E-28006-Madrid, Spain.}}}
\date{}   
\begin{document}
\maketitle
\thispagestyle{empty}
\vspace*{.5cm}

\begin{abstract}\noindent
We have computed one-loop bulk and brane mass renormalization effects
in a five-dimensional gauge theory compactified on the
$\mathcal{M}_4\times S^1/\mathbb{Z}_2$ orbifold, where an arbitrary
gauge group $\mathcal{G}$ is broken by the orbifold action to its
subgroup $\mathcal{H}$. The space-time components of the gauge boson
zero modes along the $\mathcal{H}$ generators span the gauge theory on
the orbifold fixed point branes while the zero modes of the
higher-dimensional components of the gauge bosons along the
$\mathcal{G}/\mathcal{H}$ generators play the role of Higgs fields
with respect to the gauge group $\mathcal{H}$. No quadratic
divergences in the mass renormalization of the gauge and Higgs fields
are found either in the bulk or on the branes. All brane effects for
the Higgs field masses vanish (only wave function renormalization
effects survive) while bulk effects are finite and can trigger,
depending on the fermionic content of the theory, spontaneous Hosotani
breaking of the brane gauge group $\mathcal{H}$. For the gauge fields
we do find logarithmic divergences corresponding to mass
renormalization of their heavy Kaluza-Klein modes. Two-loop brane
effects for Higgs field masses are expected from wave
function renormalization brane effects inserted into finite bulk mass
corrections.
\end{abstract}
\vspace{2.cm}   
   
\begin{flushleft}   
April 2002 \\   
\end{flushleft}
\newpage

\section{\sc Introduction}
\label{introduction}

Extra dimensions (with respect to the four space-time dimensions) are
a common ingredient in all fundamental theories aiming to unify
gravity with the rest of the known interactions. However, unlike
gravitational interactions that propagate in the bulk of the higher
dimensional space (with ten/eleven dimensions in string/M theory),
gauge interactions can propagate on a (4+d)-dimensional ($d\geq 1$)
slice of space-time, e.g. in the worldvolume of a D-brane in type
$I/I'$ strings~\cite{Polchinski:1998rr}. Moreover, it has been shown
that in these theories the radius $R$ of the compact dimensions where
the gauge interactions propagate can be large
enough~\cite{Antoniadis:1990ew} for the corresponding excitations to
be at the reach of future accelerators~\cite{Antoniadis:1993fh,
Antoniadis:1994yi, Nath:1999fs, Antoniadis:1999bq, Rizzo:1999br,
Casalbuoni:1999ns, Delgado:1999sv, Accomando:1999sj,
Antoniadis:2000vd}, while the string (or higher dimensional quantum
gravity) scale $M_s$ can be lowered to the TeV
range~\cite{Lykken:1996fj} and show up in
colliders~\cite{Giudice:1998ck, Mirabelli:1998rt} and gravitational
experiments~\cite{Adelberger:2002ic}.  This fact opened up for the
first time exciting possible experimental accessibility to fundamental
theories and provided new insight into long-standing problems of
particle physics such as the hierarchy problem.

The hierarchy problem of the Standard Model i.e. the appearance of
quadratic divergences in the quantum corrections to the Higgs mass is
one of the most outstanding problems in particle physics. It has
motivated, as the prototype perturbative solution, the introduction of
supersymmetry -- a symmetry responsible for the absence or
cancellation of quadratic divergences -- that is being looked for
extensively in experimental searches. However, in view of the
elusiveness of supersymmetry to show up in direct searches and the
robustness of the Standard Model predictions that is pushing up the
scale of supersymmetry breaking, it is interesting to explore new
avenues and possible alternative solutions to the hierarchy
problem. The existence of TeV extra dimensions where the Standard
Model fields propagate provides new and useful tools for this search.

Large extra dimensions have shown to shed new light on the hierarchy
problem. In particular, in higher-dimensional supersymmetric theories
it has been proved that one-loop radiative corrections to the Higgs
mass are finite (ultraviolet insensitive) and $\sim
1/R$~\cite{Pomarol:1998sd,Antoniadis:1998sd, Delgado:1998qr,
Barbieri:2000vh, Arkani-Hamed:2001mi, Delgado:2001si, Delgado:2001ex,
Contino:2001gz, Kim:2001re, DiClemente:2001sv}.  Of course, since the
theory is non-renormalizable, higher-loop effects introduce through
wave function renormalization a certain ultraviolet sensitivity to the
Higgs mass that can be absorbed by the renormalization group running
of the coupling constants and hence it does not make this sensitivity
explicit at low energy~\cite{Delgado:2001xr}. In this sense, the
higher dimensionality of the theory allows to improve the solution to
the hierarchy problem with respect to four dimensional supersymmetry.

However, in the presence of large compact dimensions supersymmetry is
not as necessary an ingredient as it is in four dimensions. In fact,
non-supersymmetric solutions to the hierarchy problem based on
toroidal compactifications were already explored in the
literature~\cite{Hatanaka:1998yp, Hatanaka:1999sx}. In those cases the
Standard Model Higgs field should be identified with an extra
dimensional component of a gauge field and electroweak symmetry
breaking proceeds by the Hosotani mechanism~\cite{Hosotani:1983xw,
Hosotani:1989bm}. The higher dimensional gauge invariance protects the
Higgs mass from quadratic divergences at the quantum level and
radiative corrections to the Higgs mass are finite and $\sim 1/R$. In
short, in these non-supersymmetric models the role of supersymmetry
preventing quadratic divergences is played by higher dimensional gauge
invariance.

A word of caution should be said here about the proposed
(perturbative) solutions to the hierarchy problem. All of them are
based upon introducing a symmetry at an intermediate scale $M_0$
between the electroweak scale $M_{weak}$ and the Standard Model cutoff
(quantum gravity or string scale $M_s$) such that quadratic
divergences are canceled at scales $\mu>M_0$.  In this way quadratic
divergences survive only for scales smaller than $M_0$ and radiative
corrections to the Higgs mass are $\sim M_0$. For the case of four
dimensional supersymmetry, $M_0=M_{SUSY}$, and at scales
$\mu>M_{SUSY}$ the Higgs mass is protected by supersymmetry. For the
case of a higher-dimensional non-supersymmetric theory with
electroweak Hosotani breaking, $M_0=1/R$. For scales $\mu<1/R$ the
theory is four dimensional and it is not protected from quadratic
divergences, while for $\mu>1/R$ the theory is higher-dimensional and
the gauge invariance protects the Higgs mass from quadratic
divergences. In both cases, for the mechanism to be effective, the
scale $M_0$ has to be stabilized and should be not much higher than
the electroweak scale. This comment applies to both the scale of
supersymmetry breaking $M_{SUSY}$ and to the compactification scale
$1/R$.  In particular, fixing the radius $R$ implies considering the
gravitational sector of the theory involving the radion field. This
problem is outside the scope of the present paper and we will assume
that the radius has been fixed and stabilized by some
mechanism~\cite{Ponton:2001hq}.

In the construction of higher dimensional theories the nature of the
compact space plays a prominent role in physics. In particular,
theories with more than four flat dimensions or higher dimensional
theories compactified on tori are non-chiral from the four dimensional
point of view. The simplest solution to this problem~\footnote{There
are other solutions that have been proposed involving smooth manifolds
with non-trivial backgrounds~\cite{Dvali:2001qr}.} is compactification
on tori modded out by a discrete symmetry group acting non-freely
(with fixed points) on the compact space, or orbifold
compactifications~\cite{Dixon:1985jw, Dixon:1986jc}.  Orbifolds are
not smooth manifolds but have singularities at the fixed points which
are four-dimensional hypersurfaces or boundaries of the
higher-dimensional space. Those boundaries will be (and are) often
named ``branes'' by an abuse of language. At the field theory level
brane contributions arise in the higher-dimensional Lagrangian by
means of Dirac delta functions.  Non-supersymmetric models on orbifold
compactifications were already proposed in~\cite{Antoniadis:2000tq,
Antoniadis:2001cv}.

It has been proved that under radiative corrections a theory with no
brane couplings will generally flow to one with non-trivial physics on
the brane~\cite{Georgi:2000wb, Georgi:2000ks, Goldberger:2001tn,
Contino:2001si}.  In particular, wave function renormalization effects
localized on the brane have been found. We could understand the
appearance of those renormalization effects since they are consistent
with the four dimensional nature and symmetries of the branes. In the
case at hand, we perform electroweak breaking by the Hosotani
mechanism in a higher-dimensional gauge theory broken by the orbifold
action to the Standard Model gauge theory on the brane. The Higgs is a
scalar field from the point of view of the brane and its mass is not a
priori protected by the higher-dimensional gauge invariance from
acquiring quadratic divergences localized on the brane.  Possible mass
terms on the branes that are not protected by the residual gauge
invariance are not obviously protected from the higher dimensional
symmetries either and will be investigated in detail in this
paper. Their absence should be essential for any phenomenological
applications aiming to solve the hierarchy problem in the absence of
supersymmetry.

The plan of this paper is as follows. In section~\ref{generalgg} we
will consider a five dimensional (5D) theory compactified on the
orbifold ${\cal M}_4\times S^1/\mathbb{Z}_2$ where an arbitrary gauge
group $\cal G$ is broken to the subgroup ${\cal H}$ by the orbifold
action. We also consider fermions in an arbitrary representation ${\bf
R}$ of the gauge group and the associated, consistent with the
orbifold, $\mathbb{Z}_2$ parity action. We fix the gauge consistently
with the properties of the 5D theory and introduce the corresponding
Faddeev-Popov ghosts with their transformation properties under the
orbifold action. The general Feynman rules for fields propagating in
the bulk of the orbifold are exhibited explicitly as well as some
useful group theoretical formulas that will be used in the rest of the
paper. A general discussion of the allowed orbifold gauge breaking
patterns and associated consistent fermion representations is
postponed to appendix~\ref{Appendix} while the gauge fixing conditions
and the unitary gauge are discussed in appendix~\ref{AppendixB}. In
section~\ref{one-loop} we discuss the general structure of corrections
generated in the bulk and on the orbifold fixed planes by radiative
corrections in the bulk. We will restrict ourselves to radiative
corrections to mass terms, i.e. corrections generated by diagrams with
vanishing external four-momentum: $p_\mu=0$.  In
section~\ref{gaugetwo} the one-loop radiative corrections to all field
masses are computed in the theory where an arbitrary gauge group
$\mathcal{G}$ is broken to the subgroup $\mathcal{H}$ by the orbifold
action with the fermions in an arbitrary representations ${\bf R}$ of
the gauge group.  We have considered separately gauge and fermion
sectors and bulk and brane effects. We have found no quadratic
divergences either in the bulk or on the branes. While this effect in
the bulk is justified from the higher dimensional gauge invariance,
its interpretation on the brane for the extra-dimensional components
of gauge fields is less clear, although it might be related to the
higher-dimensional Lorentz and gauge symmetries.  On the other hand,
squared mass terms are generated for the extra-dimensional components
of the gauge fields opening up the possibility of spontaneous breaking
of the residual $\mathcal{H}$ gauge symmetry on the branes.  In
particular, the contribution from the gauge (fermion) sector to the
squared mass terms is positive (negative). In section~\ref{hosotani}
the conditions for Hosotani breaking are discussed and shown to depend
on the group-theoretical invariants of the gauge group and fermion
representations. Also the possibility of reducing the rank of
$\mathcal{H}$ by the Hosotani breaking is briefly discussed. This
possibility is essential in model building where one could identify
$\mathcal{H}$ with the Standard Model gauge group.  Finally some
comments about two-loop corrections and our conclusions are drawn in
section~\ref{conclusion}.

\section{\sc Broken gauge symmetry on the $\mathbb{Z}_2$ orbifold}
\label{generalgg}

The model we will consider is a gauge theory coupled to matter in five
flat space-time dimensions with coordinates $x^M=(x^\mu,x^5)$ and
metric signature $(+,-,-,-,-)$.  The 5D geometry is ${\cal M}_4\times
S^1/\mathbb{Z}_2$, i.e. the fifth dimension is compactified on a
$\mathbb{Z}_2$ orbifold, whereas the remaining part is four
dimensional Minkowski space with metric $\eta_{\m\n}$.  We have
neglected gravity so the fixed planes having no tension are rigid
geometrical boundaries. Keeping this remark in mind we will refer to
these planes as "branes".  We denote the radius of the compact circle
by $R$.  The gauge group in the bulk is $\cal G$ (we denote $dim({\cal
G})\equiv d_{\cal G}$) and it is broken to ${\cal H}={\cal H}_1\otimes
{\cal H}_2\otimes\dots$ (we denote $dim({\cal H})\equiv d_{\cal H}$)
on the fixed hyperplanes by our choice of the orbifold projection.
For matter, we couple Dirac fermions ${\Psi_R}$ that transform in the
representation ${\bf R}$ (we denote $dim({\bf R})\equiv d_{R}$) of
$\cal G$ to the gauge fields.~\footnote{There should be no confusion
between the representation ${\bf R}$ and the radius $R$.}  We will use
capital letters from the beginning of the Latin alphabet to denote
gauge indices ($A,B,C,\cdots $), capital letters from the middle of
the Latin alphabet to denote five dimensional Lorentz indices
($M,N,R,\cdots $), small letters from the middle of the Greek alphabet
to denote four dimensional Lorentz indices ($\m,\n,\rho, \cdots $),
and small letters from the middle of the Latin alphabet to denote the
discrete fifth dimensional momentum ($k,l,m, \cdots $).
  
Our starting point is the action 
\begin{equation} S_5 = \int d^5x \Tr\left\{-\frac{1}{2}
{F}_{MN}{F}^{MN}+i{\overline \Psi_R}
\g^MD_M{\Psi_R}\right\},\label{5daction}
\end{equation}
where ${F}_{MN}=F_{MN}^AT_R^A$,
$F_{MN}^A=\partial_{M}A_N^A-\partial_{N}A_M^A +gf^{ABC}A_M^BA_N^C$
with the indices $A,B,C$ running over the adjoint representation of
the gauge group and $f^{ABC}$ the corresponding structure
constants. The gauge covariant derivative is $D_{M}=\partial
_{M}-igA_{M}^AT_R^A$, where $T_R^A$ are matrices corresponding to the
representation ${\bf R}$~\footnote{When the subscript $R$ is omitted
it is implied that the matrices are in the fundamental
representation.}  of the gauge group satisfying
\begin{equation} [T_R^A,T_R^B]=if^{ABC}T^C_R.\end{equation}
%
Our parity assignment is defined by
\begin{equation}
A_{M}^{A}(x^{\m},-x^5)=\a^M \Lambda^{AB} A_{M}^{B}(x^{\m},x^5)
\hskip .5cm (\text{no sum over } M) \label{Lambdaalpha}
\end{equation}
\begin{equation} \Psi_R(x^{\m},-x^5)=\l_R \otimes (i\g^5)\Psi_R(x^{\m},x^5), 
\label{pfermion}\end{equation} 
where $\L$ and $\l_R \otimes (i\g^5)$ represent the $\mathbb{Z}_2$
action on the gauge bosons and the fermions respectively ($\l_R$ acts
on the representation indices), $\g^5=diag(-i,i)$ and $\a^{\m}=+1$,
$\a^{5}=-1$. In addition, $\l_R$ is a hermitian matrix that squares
to one and therefore unitary.  Consistency of the 5D gauge symmetry
with the orbifold action requires the condition~\cite{Hebecker:2001jb}
\begin{equation} f^{ABC}=\L^{AA'}\L^{BB'}\L^{CC'}f^{A'B'C'},
\label{automorphism1}\end{equation} 
where summation over repeated indices is understood.  The above
constraint comes from the requirement that under the $\mathbb{Z}_2$
action $F_{MN}^A\rightarrow \a^M\L^{AB}F_{MN}^B$ (no sum over $M$), so
that $F_{MN}^AF^{AMN}$ is invariant and it is straightforward to check
that it is an automorphism of the Lie algebra of ${\cal G}$.

On the other hand, the invariance of
the fermion kinetic term requires the transformation
\begin{equation}
{\cal P}_\Psi \g^M{\cal P}_\Psi=\pm\a^M\g^M\ns
,
\label{gammatrans}
\end{equation}
while the invariance of the fermion-gauge boson coupling implies in
addition that
\begin{equation}
{\cal P}_\Psi T^A_R{\cal P}_\Psi=\L^{AB} T^B_R.
\label{Ttrans}
\end{equation}
In five dimensions the only solution to these equations is ${\cal
P}_\Psi=\l_R\otimes (i\g_5)$ (with $\lambda_R$ satisfying
Eq.~(\ref{Ttrans})), which corresponds to the lower sign in
Eq.~(\ref{gammatrans}). With no loss of generality we can diagonalize
$\L^{AA'}=\eta^A\delta^{AA'}$ with $\eta^A=\pm1$, consequently
Eq.~(\ref{automorphism1}) takes the simpler form
\begin{equation} f^{ABC}=\eta^{A}\eta^{B}\eta^{C}f^{ABC} \label{auto1}\ns.\end{equation} 
We will then often express (\ref{Lambdaalpha}) succintly
as 
\begin{equation}
A_{M}^{A}(x^{\m},-x^5)=\a^M \eta^A A_{M}^{A}(x^{\m},x^5)
{\ns}. \label{etaalpha}
\end{equation}
A few remarks are in order.
From (\ref{auto1}) we can see that since the orbifold can break ${\cal G}$ 
completely only if ${\cal G}=U(1)$, the 
unbroken subgroup is always (except in the $U(1)$ case) 
a non trivial subgroup of ${\cal G}$
\footnote{In particular, for models where the bulk gauge symmetry
coincides with the gauged superalgebra of a locally supersymmetric
theory, this means that an extended bulk supersymmetry (${\cal N}>1$)
cannot be completely broken on the fixed planes by the
$\mathbb{Z}_2$.}.  Under the $\mathbb{Z}_2$ action we can naturally
split the adjoint index $A$ into an unbroken part $a$ and a broken
part ${\hat a}$ so that the generators of the subgroup ${\cal H}$ are
$T_R^a$ and the generators of the coset ${\cal K}={{\cal G}/{\cal H}}$
(we denote $dim({\cal K})\equiv d_{\cal K}$) are $T_R^{\hat a}$.
Notice that constraint (\ref{auto1}) simply means that the matrix $\L$
is a diagonal matrix with $d_{\cal H}$ elements equal to +1 and with
the rest of the $d_{\cal K}$ elements, corresponding to the broken
part of ${\cal G}$, equal to $-1$.  According then to
Eq.~(\ref{etaalpha}), only $A_{\m}^{A}$ with $\eta^A=+1$ and
$A_{5}^{A}$ with $\eta^A=-1$ acquire zero modes and they are
non-vanishing at the fixed planes. The former appear as the (massless)
four dimensional gauge fields that correspond to ${\cal H}$ and the
latter as massless scalar fields in four dimensions.  Also
(\ref{pfermion}) results in a non trivial constraint on the possible
bulk fermion representation choices. Let us assume for now that we
have made a consistent choice for $\l_R$. Then, the fermions that will
appear massless in four dimensions will be chiral because of the
$i\g^5$ appearing in (\ref{pfermion}) and they will transform in some
representation ${\bf R}=\bigoplus_i{\bf r}_i$ of ${\cal H}={\cal
H}_1\otimes {\cal H}_2\otimes\dots$.  Another important constraint is
that the resulting massless chiral spectrum on the fixed hyperplane
should be anomaly free. This puts further restrictions on realistic
model building. There are several different ways to arrive at an
anomaly free model but this is not the subject of the present work.
In appendix~\ref{Appendix} we work out a few simple examples to
illustrate issues related to the $\mathbb{Z}_2$ action on the
fermions.

We will now use the formalism of~\cite{Georgi:2000wb, Georgi:2000ks},
i.e. we will work with exponential modes for the fields. The modes for
any field $\phi$ are related by
\begin{equation} \phi^{-m}={\cal P}_{\phi}\, \phi^m, 
\label{orbcond}\end{equation}
where ${\cal P}_{\phi}$ is the parity operator of the field
$\phi$. For fermions it is \ ${\cal P}_{{\Psi}}=\l_R \otimes (i\g^5)$
as in (\ref{pfermion}), whereas for the gauge bosons it is ${\cal
P}_{A}=\a^M \d^{M'}_{M}\L^{AB}$ with eigenvalues $\a^M\eta^A$ as in
(\ref{etaalpha}).  Eq.~(\ref{orbcond}) is automatically satisfied if
the fields are expressed by unconstrained ones:
\begin{equation}
\phi^{m}=\frac{1}{2}\left(\varphi^{m}+{\cal P}_{\phi}\, \varphi^{-m}\right)
.
\end{equation}
The gauge propagator can therefore be written as
\begin{align} 
\left<A^mA^{m^\prime}\right>=&\frac{1}{2}
\left(G^{(A)}(p_\m,p_5)
+{\cal P}_AG^{(A)}(p_\m,-p_5){\cal P}_A\right)\delta_{m-{m^\prime}}
\nonumber\\
+&\frac{1}{2}
\left({\cal P}_AG^{(A)}(p_\m,-p_5)
+G^{(A)}(p_\m,p_5){\cal P}_A\right)\delta_{m+{m^\prime}}
\label{bosonprop}
.\end{align}
In the above, we have denoted by $G^{(A)}(p_\m,p_5)$ the 5D propagator
that corresponds to the compactification on $S^1$, $p_\m$ is the
momentum along ${\cal M}_4$, $p_5=m/R$ is the momentum in the compact
direction and we have used the notation $\delta_k\equiv\delta_{k,0}$.
Using the covariance of $G^{(A)}(p_\m,p_5)$ under parity
transformations (see (\ref{gbprop}))
\begin{equation}
{\cal P}_AG^{(A)}(p_\m,-p_5){\cal P}_A=G^{(A)}(p_\m,p_5)
\end{equation}
Eq.~(\ref{bosonprop}) can be simplified to
\begin{equation}
\left<A^mA^{m^\prime}\right>=\frac{1}{2}G^{(A)}(p_\m,p_5)
\left(\delta_{m-{m^\prime}}+{\cal P}_A\delta_{m+{m^\prime}}\right)
.
\end{equation}
The fermionic propagator can be computed similarly taking into
account that ${\cal P}_\Psi \g^M{\cal P}_\Psi=-\a^M\g^M$
(no sum), which in turn implies the transformation (see (\ref{fprop}))
\begin{equation}
{\cal P}_\Psi G^{(\Psi)}(p_\m,-p_5){\cal P}_\Psi=-G^{(\Psi)}(p_\m,p_5).
\end{equation}
Then the simplified fermion propagator reads
\begin{equation}
\left<\Psi^m\overline{\Psi}^{m^\prime}	\right>=\frac{1}{2}G^{(\Psi)}(p_\m,p_5)
\left(\delta_{m-{m^\prime}}-{\cal P}_\Psi\delta_{m+{m^\prime}}\right)
.
\end{equation}
The vertices then conserve 5D momentum and are the ones of the
unorbifolded 5D theory. All the information about the non-trivial
$\mathbb{Z}_2$ action is encoded in the propagators in a particularly
simple way. 

The next issue is gauge fixing and ghosts.  We will work in the 5D
covariant gauge $\partial_M\,A^M=0$.  The modified Lagrangian then
including the gauge fixing term and the ghost fields $c^A$ is in the
standard way
\begin{equation} {\cal L}_5\rightarrow {\cal L}_5-\frac{1}{2\xi}
\partial^{M}A_{M}^B\partial^{N}A_{N}^B
+\Tr{\partial^{M}{\overline c}D_{M}c}.\end{equation}
By looking at the ghost-gauge field interaction term above,
we can see that the ghost $c^B$  has the same parity as the
gauge field with the same index $B$, i.e.
\begin{equation}
c^{A}(x^{\m},-x^5)=\Lambda^{AB}
c^{B}(x^{\m},x^5)
\label{pghost}
\end{equation}
The propagators and vertices can then be taken over from any standard
textbook with the indices properly generalized to five dimensions. For
completeness we give their explicit forms in 5D Minkowski
space-time~\cite{Peskin:1995ev}:
\begin{equation} 
G^{(A)}(p_\m,p_5)= \quad
\begin{picture}(60,40)(0,17)
\Photon(0,20)(60,20)25
\footnotesize
\Text(0,30)[l]{$B,M$}
\Text(60,30)[r]{$C,N$}
\LongArrow(22,15)(38,15)
\Text(30,10)[c]{$(p,p_5)$}
\end{picture}
\quad
=-i\frac{\d^{BC}}{p^2-{p_5}^2}
\left(g_{MN}-(1-\xi)\frac{p_Mp_N}{p^2-{p_5}^2}\right)
\label{gbprop}
\end{equation}
\be G^{(c)}(p_\m,p_5)=
\quad
\begin{picture}(60,40)(0,17)
\DashLine(0,20)(60,20)4
\footnotesize
\Text(0,30)[l]{$B,M$}
\Text(60,30)[r]{$C,N$}
\LongArrow(22,15)(38,15)
\Text(30,10)[c]{$(p,p_5)$}
\end{picture}
\quad
=
i\frac{\d^{BC}}{p^2-{p_5}^2}
\label{ghostprop}
\end{equation}
\begin{equation} 
G^{(\Psi)}(p_\m,p_5)=\quad
\begin{picture}(60,40)(0,17)
\Line(0,20)(60,20)
\footnotesize
\Text(0,30)[l]{$a$}
\Text(60,30)[r]{$b$}
\LongArrow(22,15)(38,15)
\Text(30,10)[c]{$(p,p_5)$}
\end{picture}
\quad
=i\frac{{\d^a}_b}{{\g^{\m}p_{\m}+\g^5{p_5}}}
\label{fprop}
\end{equation}
\vspace{-1.2cm}
\brr 
\begin{picture}(80,60)(-20,27)
\Photon(5,5)(30,30)23
\Photon(5,55)(30,30)23
\Photon(30,30)(60,30)23
\footnotesize
\LongArrow(6,12)(16,22)
\Text(7,17)[r]{$(r,r_5)$}
\Text(7,43)[r]{$(q,q_5)$}
\Text(45,43)[c]{$(p,p_5)$}
\LongArrow(6,48)(16,38)
\LongArrow(52,35)(38,35)
\Text(0,0)[c]{$C,R$}
\Text(0,60)[c]{$B,N$}
\Text(60,20)[r]{$A,M$}
\end{picture}
=g\ f^{ABC}\bigl((r-q)_{M}g_{NR}+
(q-p)_{R}g_{MN}+(p-r)_Ng_{RM}\bigr)\nonumber\\
{\rm with} \hskip 1cm p+q+r=0 
\label{ggg}
\err
\vspace{-1.2cm}
\begin{align} 
\begin{picture}(100,70)(-20,27)
\Photon(5,5)(55,55)26
\Photon(5,55)(55,5)26
\footnotesize
\LongArrow(6,12)(16,22)
\Text(7,17)[r]{$(r,r_5)$}
\Text(7,43)[r]{$(q,q_5)$}
\Text(55,17)[l]{$(s,s_5)$}
\Text(55,43)[l]{$(p,p_5)$}
\LongArrow(6,48)(16,38)
\LongArrow(54,48)(44,38)
\LongArrow(54,12)(44,22)
\Text(0,0)[c]{$C,R$}
\Text(0,60)[c]{$B,N$}
\Text(55,0)[c]{$D,S$}
\Text(55,60)[c]{$A,M$}
\end{picture}
=
-ig^2\ \bigl(&f^{ABE}f^{CDE}
(g_{MR}g_{NS}-g_{NR}g_{MS})\nonumber\\
+&f^{CBE}f^{ADE}
(g_{MR}g_{NS}-g_{NM}g_{RS})\nonumber\\ 
+&f^{DBE}f^{CAE}
(g_{SR}g_{NM}-g_{NR}g_{MS})\bigr)
\label{gggg}
\end{align}
\vspace{-1cm}
\begin{equation} 
\begin{picture}(80,60)(-20,27)
\Line(5,5)(30,30)
\Line(5,55)(30,30)
\Photon(30,30)(60,30)23
\footnotesize
\LongArrow(6,12)(16,22)
\Text(7,17)[r]{$(r,r_5)$}
\Text(7,43)[r]{$(q,q_5)$}
\Text(45,43)[c]{$(p,p_5)$}
\LongArrow(6,48)(16,38)
\LongArrow(52,35)(38,35)
\Text(0,0)[c]{$c$}
\Text(0,60)[c]{$b$}
\Text(60,20)[r]{$A,M$}
\end{picture}
=ig\ \g_M{(T^A)^c}_b
\label{fgg}
\end{equation}
\vskip .4cm
\begin{equation} 
\begin{picture}(80,60)(-20,27)
\DashLine(5,5)(30,30)4
\DashLine(5,55)(30,30)4
\Photon(30,30)(60,30)23
\footnotesize
\LongArrow(6,12)(16,22)
\Text(7,17)[r]{$(r,r_5)$}
\Text(7,43)[r]{$(q,q_5)$}
\Text(45,43)[c]{$(p,p_5)$}
\LongArrow(6,48)(16,38)
\LongArrow(52,35)(38,35)
\Text(0,0)[c]{$C$}
\Text(0,60)[c]{$B$}
\Text(60,20)[r]{$A,M$}
\end{picture}
=g\ f^{ABC}q_M
\label{ccg}
\end{equation}
\vspace{1cm}

Finally, in our loop computations, we will 
make use of the following formulas:
\begin{eqnarray}
f^{ABC}f^{A'BC}&=&C_2({\cal G})\delta^{AA'}\label{identitiesfirst}\\
f^{ABC}f^{A'BC}\eta^C&=&\bigl(C_2({\cal H}_A)- 
\frac{1}{2}C_2({\cal G})\bigr)(\eta^A+1)\delta^{AA'}\\
f^{ABC}f^{A'BC}\eta^B\eta^C&=&C_2({\cal G})\eta^A\delta^{AA'}\\
\text{tr}~(T_R^A T_R^{A'})&=&C_R\delta^{AA'}\\
\text{tr}~(T_R^A\l_R T_R^{A'}\l_R)&=&C_R\eta^A\delta^{AA'},
\label{identitieslast}\end{eqnarray}
where $C_2(\mathcal{G})$ is the quadratic Casimir of ${\cal G}$ and
$C_R$ is the Dynkin index of the representation ${\bf R}$ satisfying
$d_RC_2(R)=C_Rd_{\cal G}$.  We normalize the fundamental
representation to have its Dynkin index equal to 1/2. In the second
identity we have called ${\cal H}_A$ the unbroken subgroup which
$T_R^A$ belongs to~\footnote{Since in the subspace of unbroken
generators $\Lambda$ is just the identity one can find a basis in
which the groups ${\cal H}_i$ are generated by some
$\{T^{A_i}\}$.}. Note that this makes sense since the whole expression
vanishes if $T_R^A$ is a broken generator $T_R^A=T_R^{\hat a}$, (i.e.\
when $\eta^A=-1$).

We will find it convenient to work in the Feynman gauge, $\xi=1$. For
any other value of $\xi$, including the 5D Landau gauge $\xi=0$, there
is a tree-level mixing $\langle A_\mu^A A_5^B \rangle$ as can be seen
from Eq.~(\ref{gbprop}). Furthermore in the class of gauges defined by
$\partial_M A^M=0$ there is no value of $\xi$ for which $\langle
A_5^{A,n} A_5^{B,n} \rangle=0$, $n\neq 0$, which would correspond to
the unitary gauge since the fields $A_5^{A,n}$ ($n\neq 0$) are the
Goldstone bosons corresponding to the orbifold gauge breaking. To
reach the unitary gauge the gauge fixing condition has to be defined
in a 5D non-covariant fashion, just consistent with the compactified
theory. This issue was discussed in Ref.~\cite{Papavassiliou:2001be}
for an abelian gauge theory. It is discussed in
appendix~\ref{AppendixB} for a general non-abelian group broken to a
subgroup by the orbifold compactification.

\section{\sc The structure of loop corrections}
\label{one-loop}

In this section we will analyze what kind of terms in the effective
action could be generated in the bulk and on the orbifold fixed planes
by radiative corrections in the bulk.  We will focus on the bilinear
terms in the effective action and also look only at terms not
involving derivatives with respect to infinite directions,
$\partial^\mu$. The most general terms are then:
\begin{equation}
\Gamma[A]=\int d^5x \left\{A_M^B\Pi_{MN}^{BC}[-\partial_5^2]A_N^C+
\frac{1}{2}\bigl(\delta(x_5)+\delta(x_5-\pi R)\bigr)
A_M^B\widetilde\Pi_{MN}^{BC}
[i\!\stackrel{\leftarrow}{\partial_5}, i\!\stackrel{\rightarrow}{\partial_5}]
A_N^C\right\}
\label{effact}
.
\end{equation}
The arrows on the $\partial_5$ indicate the field whose derivative is
to be taken.  The bulk and brane terms in Eq.~(\ref{effact}) can be
read off from the generic two-point function 
\vspace{-1.5cm}
\begin{equation}
\left<A_M^{B,m}A_N^{C,m'}\right>=
\ \begin{picture}(120,60)(0,18)
\GCirc(60,20){15}{0.7}
\Photon(0,20)(45,20)24 
\Photon(75,20)(120,20)24 
\footnotesize
\Text(0,12)[l]{$B,M$}
\Text(0,28)[l]{$m$}
\Text(120,12)[r]{$C,N$}
\Text(120,28)[r]{${m^\prime}$}
\end{picture}
\label{generic}
\end{equation} 
\vskip .3cm
\noindent 
in the following way: The contributions to Eq.~(\ref{generic}) for
which the five momentum in the outgoing line is conserved, $m'=\pm m$,
will give rise to bulk terms. We will denote them as
\begin{equation}
\langle A_M^{B,m}A_N^{C,m'}\rangle_{\rm bulk}=\Pi_{MN}^{+BC}(m^2)\d_{m-m'}
	+\Pi_{MN}^{-BC}(m^2)\d_{m+m'}.
\label{notation1}
\end{equation}
Contracting this expression with the external fields one finds
\begin{equation}
\sum_{m,m'}A_M^{B,m}\left(\Pi_{MN}^{+BC}(m^2)\d_{m-m'}
	+\Pi_{MN}^{-BC}(m^2)\d_{m+m'}\right)A_N^{C,m'}.
\end{equation}
We now take the Fourier transform of Eq.~(\ref{effact}),
sum over $m'$: 
\begin{equation}
\int dx_5\ A_M^B\Pi_{MN}^{BC}[-\partial_5^2]A_N^C=
	\sum_{m,m'} A_M^{B,m}\Pi_{MN}^{BC}(m^2)\d_{m-m'}
	A_N^{C,m'}
\end{equation}
and read off the correspondence:
\begin{equation}
\Pi_{MN}^{BC}(m^2)=\Pi_{MN}^{+BC}(m^2)+\alpha^N\eta^C\Pi_{MN}^{-BC}(m^2).
\label{SumPi}
\end{equation}
In the second term after summing over $m'$ there appears a factor of
$A_N^{C,-m}$ which has to be transformed into $A_N^{C,+m}$ at the cost
of an $\a^M\eta^C$, according to (\ref{orbcond}).  By doing a Taylor
expansion around $m=0$ we obtain the possible bilinear operators in
the bulk:
\begin{equation} A^m \Pi(m^2) A^{m}=
A^{m}\Pi^{(0)}A^{m}+\frac{1}{2}A^{m}\Pi^{(2)}m^2A^{m}+
\frac{1}{24}A^{m}\Pi^{(4)}m^4A^{m}+\dots.
\label{bulkexpansion}
\end{equation}
The summation over $m$ and the gauge and Lorentz indices are left
implicit and $\Pi^{(n)}$ is the $n$th derivative with respect to $m$
evaluated at $m=0$.  The bulk mass term is just given by $\Pi^{(0)}$
($=\Pi{(0)}$).

The terms in Eq.~(\ref{effact}) proportional to $\delta$-functions
are generated by contributions in Eq.~(\ref{generic}) which do not
conserve five-momentum.  Instead we will find expressions of the form
\begin{equation}
\langle A_M^{B,m}A_N^{C,m'}\rangle_{\rm brane}=\sum_l\biggl\{
	\widetilde\Pi_{MN}^{+BC}(m,l)\d_{m-m'-2l}+
	\widetilde\Pi_{MN}^{-BC}(m,l)\d_{m+m'-2l}
	\biggr\}.
\label{notation2}
\end{equation}
Note that the sum over $l$ is now constrained through the Kronecker-$\d$,
yielding an amplitude which depends on the two independent variables
$m$ and $m'$.  To identify $\widetilde\Pi_{MN}^{BC}(m,m')$ we contract
this matrix with $A_M^{B,m}$ and $A_N^{C,m'}$ and perform the sum over
$l$:
\begin{equation}
\sum_{m,m';\,m\pm m'{\rm even}}A_M^{B,m}\biggl\{
	\widetilde\Pi_{MN}^{+BC}(m,\frac{m-m'}{2})+
	\widetilde\Pi_{MN}^{-BC}(m,\frac{m+m'}{2})
	\biggr\}A_N^{C,m'}.\label{braneeffterm}
\end{equation}
To compare this with Eq.~(\ref{effact}) one takes the Fourier
transform of the latter with respect to the compact dimension:
\begin{align}
\int dx_5\ 
\frac{1}{2}\bigl(\delta(x_5)&+\delta(x_5-\pi R)\bigr)
A_M^B\widetilde\Pi_{MN}^{BC}
\bigl[i\!\stackrel{\leftarrow}{\partial_5}, i\!\stackrel{\rightarrow}
{\partial_5}\bigr]
A_N^C\nonumber\\
&=\sum_{m,m',k}\int dx_5\ \frac{1}{2}e^{-i(m+m'+k)x_5}(1+e^{ik\pi})A_M^{B,m}
\widetilde\Pi_{MN}^{BC}(m,m')
A_N^{C,m'}\nonumber\\
&=\sum_{m,m';\, m\pm m'{\rm even}}A_M^{B,m}
\widetilde\Pi_{MN}^{BC}(m,m')
A_N^{C,m'}.
\end{align}
We conclude that 
\begin{equation}
\widetilde\Pi_{MN}^{BC}(m,m')=
\widetilde\Pi_{MN}^{+BC}(m,\frac{m-m'}{2})+\widetilde
\Pi_{MN}^{-BC}(m,\frac{m+m'}{2})
.\label{SumPitilde}
\end{equation}
To understand the brane terms a bit better, one can make a Taylor
expansion of (\ref{braneeffterm}) around $m,m'=0$. One gets for even
($E$) and odd ($O$) fields respectively:
\begin{align}
E^m
\widetilde\Pi(m,m')
E^{m'}=&E^{m}\widetilde\Pi^{(0,0)}E^{m'}+
\frac{1}{2}(m^2E^{m})\widetilde\Pi^{(2,0)}E^{m'}+
\frac{1}{2}E^{m}\widetilde\Pi^{(0,2)}(m^{\prime 2} E^{m'})+\dots,\nonumber\\
O^m
\widetilde\Pi(m,m\prime)
O^{m\prime}=&(mO^{m})\widetilde\Pi^{(1,1)}(m\prime O^{m\prime})+
\frac{1}{6}(mO^{m})\widetilde\Pi^{(1,3)}(m^{\prime 3} O^{m'})+\dots,
\label{braneexpansion}
\end{align}
where we have suppressed the gauge and Lorentz indices for clarity and
the (independent) summations over $m,m'$ are implicit. In particular
it is now clear that the brane mass term for even fields is just
$\widetilde\Pi^{(0,0)}$ ($=\widetilde\Pi{(0,0)}$), given by setting
$m=m'=0$ in the momentum-violating terms of Eq.~(\ref{generic}).  In
contrast to Eq.~(\ref{bulkexpansion}) the expansion coefficients are
now not diagonal but democratic matrices in mode space.

Bulk terms are not expected to generate any divergences that were not
present in the original 5D theory.  Brane terms, however, could
generate new divergences. In particular, we want to investigate the
possible appearance of a scalar mass on the brane. Had we introduced
fundamental massless scalars in the 5D theory, their masses would pick
up corrections proportional to the cutoff of the five dimensional
theory and the same would happen with their zero modes on the branes
(provided they survive the orbifold projection).  In our model there
are no scalars in the five dimensional theory. The only bosonic fields
are the components of the bulk gauge fields whose masses are zero at
tree level and remain zero at all orders because they are protected by
gauge invariance.  On the branes, however, there are scalars. Some of
them are massless at tree level, namely the zero modes of positive
$\mathbb{Z}_2$ parity fields.  The natural question that arises then
is whether these masses are protected against radiative corrections.
We know that on general grounds this is not the case and in order to
avoid the Higgs mass to pick up corrections proportional to the
cutoff, it has to be protected by some symmetry, for example
supersymmetry. Our model is not supersymmetric and therefore such a
mechanism is not possible.  On the other hand since the scalars are
extra dimensional components of the original gauge field, one would
hope that gauge invariance still protects them. The problem is that it
is on the branes where the zero modes of positive $\mathbb{Z}_2$
parity fields are seen as massless scalar fields and we have also seen
that it is on the branes that the only surviving symmetry is ${\cal
H}$; there is no apparent symmetry that prohibits dangerous
corrections to their masses. In the following we will see in a
one-loop calculation that even though there is no such apparent
symmetry on the branes, the larger, original bulk gauge symmetry
arranges the couplings of the zero modes and the Kaluza-Klein (KK)
towers in such a way that the masses of the zero modes remain
protected.

\section{\sc One loop corrections}
\label{gaugetwo}

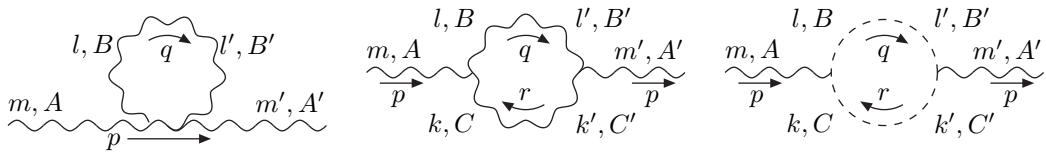
\begin{figure}
\begin{center}
\begin{picture}(120,60)(0,0)
\Photon(0,0)(120,0)29
\PhotonArc(60,20)(20,-90,250)29 	
\LongArrowArcn(60,20)(14,120,50)
\footnotesize
\Text(60,27)[c]{$q$}
\Text(0,8)[l]{$m,A$}
\Text(120,8)[r]{${m^\prime},A'$}
\Text(40,30)[r]{$l,B$}
\Text(80,30)[l]{$l',B'$}
\LongArrow(45,-5)(75,-5)
\Text(40,-6)[c]{$p$}
\end{picture}
\quad 
\begin{picture}(120,60)(0,0)
\PhotonArc(60,20)(20,0,360)29
\Photon(0,20)(40,20)23 
\Photon(80,20)(120,20)23	
\LongArrowArcn(60,20)(14,120,50)
\LongArrowArcn(60,20)(14,300,230)
\footnotesize
\Text(60,12)[c]{$r$}
\Text(60,27)[c]{$q$}
\Text(0,28)[l]{$m,A$}
\Text(120,28)[r]{${m^\prime},A'$}
\Text(41,0)[r]{$k,C$}
\Text(41,40)[r]{$l,B$}
\Text(79,0)[l]{$k',C'$}
\Text(79,40)[l]{$l',B'$}
\LongArrow(5,16)(20,16)
\Text(12,10)[c]{$p$}
\LongArrow(100,16)(115,16)
\Text(108,10)[c]{$p$}
\end{picture}
\quad 
\begin{picture}(120,60)(0,0)
\DashCArc(60,20)(20,0,360)3
\Photon(0,20)(40,20)23 
\Photon(80,20)(120,20)23 
\LongArrowArcn(60,20)(14,120,50)
\LongArrowArcn(60,20)(14,300,230)
\footnotesize
\Text(60,12)[c]{$r$}
\Text(60,27)[c]{$q$}
\Text(0,28)[l]{$m,A$}
\Text(120,28)[r]{${m^\prime},A'$}
\Text(41,0)[r]{$k,C$}
\Text(41,40)[r]{$l,B$}
\Text(79,0)[l]{$k',C'$}
\Text(79,40)[l]{$l',B'$}
\LongArrow(5,16)(20,16)
\Text(12,10)[c]{$p$}
\LongArrow(100,16)(115,16)
\Text(108,10)[c]{$p$}
\end{picture}
\end{center}
\caption{The diagrams contributing from the gauge sector}
\label{gaugediagrams}
\end{figure}
The one-loop corrections in the effective action coming from the
exchange of 5D gauge fields are given in Figs.~\ref{gaugediagrams} and
\ref{fermiondiagram}. We first discuss the general structure of the
brane and bulk terms. As a first step, we will not compute the exact
values of the diagrams. It is enough to observe their general
structure to deduce whether and when there appear bulk or brane
terms. Then, once we have separated bulk from brane terms, we carry
out the actual calculation for each case separately and interpret the
results.

Since we are mainly interested in mass corrections we will explicitly
evaluate the one-loop graphs only for vanishing external four-momentum
($p^\mu=0$). The amplitudes corresponding to each graph will be
denoted as in Eqs.~(\ref{notation1}) and (\ref{notation2}). In
addition, we add a superscript $(i)$, i.e.
\begin{equation}
\Pi_{MM'}^{(i)_{\pm}AA'}(m^2),\quad \widetilde\Pi_{MM'}^{(i)_{\pm}AA'}(m,l),
\end{equation} 
where $i=1$ for the tadpole, $i=2$ for the gauge loop, $i=3$ for the
ghost loop and $i=4$ for the fermion loop.  The other indices are as
described in the beginning of section~\ref{generalgg}.  All
computations in this section will be done in Euclidean space
\cite{Ramond:1989yd}.

\subsection{\sc The gauge sector}
\label{gaugesect}

In the sector where 5D gauge bosons are exchanged as internal lines there are
the three different Feynman diagrams shown in \fig{gaugediagrams}. All our
computations will be carried out in the $\xi =1$ (Feynman$-$'t Hooft)
gauge and dimensional regularization (with renormalization scale $\mu$) will
be used to handle divergent diagrams. Let us examine each case in turn.

\subsubsection{\sc The tadpole}
\label{thetadpole}

The tadpole is the first graph appearing in \fig{gaugediagrams}. 
It is proportional to
\begin{equation}
\delta_{m-l+l'-{m^\prime}}(\delta_{l-l'}+\a^M\eta^B\delta_{l+l'})
\delta^{BB'}f^{ABC}f^{A'B'C},\end{equation}
which, using the identities~(\ref{identitiesfirst})
$-$(\ref{identitieslast}), can be written as 
\begin{equation} C_2({\cal G})\delta_{m-{m^\prime}}\delta^{AA'}+\a^M\bigl(C_2({\cal H}_A)-
\frac{1}{2}C_2({\cal G})\bigr)(\eta^A+1)\delta_{m-{m^\prime}-2l}\delta^{AA'}.
\end{equation}
Notice that there is no term proportional to $\delta_{m+m'}$ and so no
contributions to the two-point function $\Pi_{MM'}^{(1)_-AA'}$ will
appear. The first term in the above is five momentum conserving and
thus gives only a bulk term. Its dependence on the gauge-index $A$ is
trivial and the contribution is the same for all $A_M^A$.  The second
term is momentum non-conserving and gives rise to a brane term.  Since
it contains the factor $(\eta^A+1)$ it is zero if $\eta^A=-1$. Thus
there are only brane terms if the external lines correspond to a field
$A_M^a$ where $a$ is an index of the unbroken group $\cal H$.

Applying the Feynman rules, in the bulk we obtain for either $\eta^A=\pm 1$
\begin{equation} 
\Pi_{MM'}^{(1)_+AA'}=
	-\frac{d}{2}g^2\d_{M{M'}}\d^{A{A'}}
C_2({\cal G})\mu^{4-d}\int \frac{d^{d }q}{{(2\pi)^{d}}}
\sum_{l=-\infty}^{+\infty}\frac{1} 
{{q^2+\frac{l^2}{{R^2}}}}\label{fig11}.
\end{equation}
Here we have defined $d=\delta_{\mu\mu}$.  This expression is
obviously the same along ${\cal M}_4$ or $S^1/\mathbb{Z}_2$.

The contribution of the tadpole to brane localized terms is nonzero
only for $\eta^A=+1$ and is equal to
\begin{align}
{\widetilde \Pi}_{MM'}^{(1)_+AA'}(m,l)=&
-\frac{(d-1-\a^M)}{4}g^2\d_{M{M'}}\d^{AA'}\nonumber\\
(\eta^A+1)&
\bigl(2C_2({\cal H}_A)-C_2({\cal G})\bigr)
\m^{4-d}\int \frac{d^{d }q}{{(2\pi)^{d}}}
\frac{1}{{q^2+{\frac{l^2}{R^2}}}}.
\label{tadpole}
\end{align}
Observe that there is no sum over the loop fifth momentum.

\subsubsection{\sc The gauge and the ghost loop}
\label{thegauge}

The second diagram in \fig{gaugediagrams} is proportional to 
\begin{equation}
\delta_{m-l-k}\delta_{l'+k'-{m^\prime}}
(\delta_{l-l'}+\a^N\eta^B\delta_{l+l'})\d_{NN'}\delta^{BB'}
(\delta_{k-k'}+\a^R\eta^C\delta_{k+k'})\d_{RR'}
\d^{CC'}f^{ABC}f^{A'B'C'},\end{equation}
which gives rise to the four terms
\begin{equation} C_2({\cal G})\delta_{m-{m^\prime}}
\delta^{AA'}\d_{NN'}\d_{RR'},\label{gt1}\end{equation}
\begin{equation} \a^N\a^R\eta^AC_2({\cal G})\delta_{m+{m^\prime}}
\delta^{AA'}\d_{NN'}\d_{RR'},\label{gt2}\end{equation}
\begin{equation} \a^R(\eta^A+1)\bigl(C_2({\cal H}_A)-\frac{1}{2}C_2({\cal
G})\bigr)\delta_{m+{m^\prime}-2l}
\delta^{AA'}\d_{NN'}\d_{RR'},\label{gt3}\end{equation}
\begin{equation} \a^N(\eta^A+1)\bigl(C_2({\cal H}_A)-\frac{1}{2}C_2({\cal
G})\bigr)\delta_{m-{m^\prime}-2l}
\delta^{AA'}\d_{NN'}\d_{RR'}.\label{gt4}\end{equation}
As in the case of the tadpole, there are bulk terms for both
$\eta^A=\pm 1$.  However, once again, one finds brane terms only for
$\eta^A=+1$.  

The structure for the ghost diagram is slightly simpler
since the internal propagators do not carry vector indices. The
conclusion is unmodified, brane terms are proportional to
$(\eta^A+1)$.

The diagrams with momentum conserving external lines  
corresponding to (\ref{gt1}) and (\ref{gt2}) and the analogous 
ghost diagrams give, for $p_\mu=0$,
\brr
\Pi_{M{M^\prime}}^{(i)_+ AA'}(m^2)&=&
	\frac{1}{8}g^2\delta^{AA'}C_2({\cal G})\mu^{4-d}
	\int \frac{d^{d }q}{{(2\pi)^{d}}}\sum_{l=-\infty}^{+\infty}
	\frac{N^{(i)}_{M{M^\prime}}}
	{{(q^2+\frac{l^2}{R^2})(q^2+\frac{(m-l)^2}{R^2})}},\label{fig12}\\
\Pi_{M{M^\prime}}^{(i)_- AA'}(m^2)&=&
	\alpha^{M'}\eta^{A'}\Pi_{M{M^\prime}}^{(i)_+ AA'}
\label{fig12two}
\err
where here of course $i=2,3$ only.  For the gauge loop the numerators
are given by
\begin{align}
N^{(2)}_{\mu\mu'}
	&=\delta_{\mu\mu'}\biggl(\frac{2l^2+5m^2-2ml}{R^2}+
2(3-\frac{1}{d})q^2\biggr),\\
N^{(2)}_{55}&=2q^2+\frac{d(m-2l)^2}{R^2},
\end{align}
while the ghost loop gives
\begin{equation}
N^{(3)}_{M{M'}}=2q_M(p-q)_{M'}.
\end{equation}
Evaluated explicitly this becomes
\begin{align}
N^{(3)}_{\mu\mu'}&=
-\frac{2}{d}q^2\delta_{\mu\mu'},\\
N^{(3)}_{55}&=\frac{2l(m-l)}{R^2}.
\end{align}
Obviously all contributions to $\Pi^{(i)AA'}_{MM'}$ are diagonal, in
particular there is no mixing between $M=\mu$ and $M=5$.  In deriving
the above one should not forget that the fifth component of the
momenta entering in the Feynman rules for the vertices may be flipped
due to the $\delta_{l+l'}$, etc.\ in the propagators.

Brane terms can be computed as well.  We will again perform the
calculation for $p_{\m}=0$. We find
\begin{align} 
{\widetilde \Pi}_{MM'}^{(i)_{\pm}AA'}(m,l)&=
\frac{1}{16}g^2\delta^{AA'}\nonumber\\
(\eta^A+1)&\bigl(2C_2({\cal H}_A)-C_2({\cal
G})\bigr)\mu^{4-d}\int \frac{d^{d }q}{{(2\pi)^{d}}}
\frac{{\widetilde N}^{(i)_{\pm}}_{MM'}}
{{(q^2+\frac{l^2}{R^2})(q^2+\frac{(m-l)}{R^2})}},
\label{gaugeandghost}
\end{align}
where
\begin{align}
{{\widetilde N}^{(2)_{\pm}}_{\mu\mu'}}&=-
	\frac{6l^2+3m^2-6ml}{R^2}-\frac{2}{d}(5-3d)q^2\delta_{\mu\mu'},
	\label{Ntilde2mu}\\
{{\widetilde N}^{(2)_{\pm}}_{55}}&=\pm
	d\frac{m(m-2l)}{R^2},\label{Ntilde25}\\
{{\widetilde N}^{(3)_{\pm}}_{\mu\mu'}}&=-
	\frac{2}{d}q^2\delta_{\mu\mu'},\label{Ntilde3mu}\\
{{\widetilde N}^{(3)_{\pm}}_{55}}&=\pm 2
\frac{l(m-l)}{R^2}.\label{Ntilde35}
\end{align}

\subsubsection{\sc Bulk effects from the gauge sector}
\label{bulkeffects}

In this section we will compute the bulk effects from the gauge sector
obtained in sections~\ref{thetadpole} and \ref{thegauge}.  It is
generally known from finite temperature field theory that by
compactifying on a circle no new divergences appear. We can give some
reasoning on why this should not happen by looking at a general
one-loop amplitude
\begin{equation}
\int \frac{d^{d }q}{{(2\pi)^{d}}}
\frac{1}{R}\sum_{l=-\infty}^{+\infty}g(q_\mu,l/R).
\label{general amplitude}
\end{equation}
We can perform a Poisson re-summation 
\begin{equation}
\frac{1}{R}\sum_l g(q_\mu,l/R)=\sum_k\tilde g(q_\mu,2\pi k R),
\end{equation}
where $\tilde g$ is the Fourier transform of $g$ with respect to
$q_5=l/R$. This allows us to rewrite the amplitude as
\begin{equation}
\int \frac{d^{d }q}{{(2\pi)^{d}}}
	\int\frac{dq_5}{2\pi} g(q_\mu,q_5)+
	\int \frac{d^{d }q}{{(2\pi)^{d}}}\sum_{k\neq0} 
	\int\frac{dq_5}{2\pi}e^{i (2\pi kR) q_5}g(q_\mu,q_5).
\end{equation}
Here we have extracted the term in the sum corresponding to $k=0$
which is just the five dimensional amplitude. The remaining terms
summed over give typically an exponentially suppressed function of the
four momentum squared~\footnote{The separation of the amplitude into a
five dimensional part and a finite part is similar to the approach
of~\cite{GrootNibbelink:2001bx}.}.  We will see explicit examples
below.

To collect the total bulk contribution from the gauge sector,
we have to add the different terms according to Eq.~(\ref{SumPi}):
\begin{equation}
\Pi_{MM'}^{AA'}(m^2)=\sum_{i=1}^{3}\Bigl(\Pi_{MM'}^{(i)+AA'}(m^2)+
\alpha^{M'}\eta^{A'}\Pi_{MM'}^{(i)-AA'}(m^2)\Bigr).\label{one}
\end{equation}
Concentrating first on the scalar sector, 
we find from Eqs.~(\ref{fig11}), (\ref{fig12}) and (\ref{fig12two})
\begin{equation}
\Pi_{55}^{AA'}(m^2)=\frac{1}{8}g^2\delta^{AA'}C_2({\cal G})
\mu^{4-d}\int \frac{d^{d }q}{{(2\pi)^{d}}}
\sum_{l=-\infty}^{+\infty}\left(\frac{-4d}{q^2+\frac{l^2}{{R^2}}}+2
\frac{2q^2+\frac{d(m-2l)^2+2l(m-l)}{R^2}}
{(q^2+\frac{l^2}{R^2})(q^2+\frac{(m-l)^2}{{R^2}})}
\right).
\end{equation}
Notice that the signs $\eta$ and $\alpha$ in Eq.~(\ref{fig12two})
exactly cancel the ones in Eq.~(\ref{one}) yielding a global factor of
2 in the second term.  Decomposing into partial fractions we obtain
\begin{multline}
\Pi_{55}^{AA'}(m^2)=\frac{1}{8}g^2\delta^{AA'}
C_2({\cal G})\mu^{4-d}\int \frac{d^{d }q}{{(2\pi)^{d}}}
\sum_{l=-\infty}^{+\infty}\\
\left(-2d\frac{1}{q^2+\frac{l^2}{{R^2}}}+2d\frac{1}{ 
{q^2+\frac{(m-l)^2}{{R^2}}}}
-4(d-1)
\frac{q^2+ \frac{l(m-l)}{R^2}} { 
(q^2+\frac{l^2}{R^2})(q^2+\frac{(m-l)^2}{R^2})}
\right)\label{three}.
\end{multline}
The first observation is that one can shift the summation index $l$ in
the second term by $m$ so that it cancels against the first term.  To
interpret the remaining term, let us do the integral first.  Naive
power counting indicates that a quadratic divergence and a logarithmic
divergence will appear in the result. Recall that in dimensional
regularization the appearance of quadratic divergences in $d=4$ are
signaled by poles in $d=2$ and notice that the usual factor of $d-2$
multiplying the pole that appears in conventional gauge theories is
missing. Despite this fact we will now show that all divergences
(quadratic and logarithmic) are actually absent.  According to
Eq.~(\ref{bulkexpansion}), we can extract the bulk mass term from
Eq.~(\ref{three}) by evaluating it at $m=0$:
\begin{equation}
\Pi_{55}^{AA'}(0)=-\frac{1}{8}g^2\delta^{AA'}C_2({\cal G})
\mu^{4-d}\int \frac{d^{d }q}{{(2\pi)^{d}}}
\sum_{l=-\infty}^{+\infty}\left(4(d-1)
\frac{q^2-\frac{l^2}{R^2}} 
{\left(q^2+\frac{l^2}{R^2}\right)^2}\right).
\label{two}
\end{equation}
There seems to be a quadratically divergent piece left over.  This is
not unexpected at this stage since the four dimensional gauge
invariance alone does not protect these scalars from acquiring
divergences. However, five-dimensional gauge-invariance does:
According to our discussion below Eq.~(\ref{general amplitude}) one
can extract the five dimensional part of the amplitude by substituting
the summation by an integration over $q_5=l/R$.  In this case,
however, this integral turns out to be zero and therefore quadratic
divergences are absent.  The remaining terms of the Poisson
re-summation (sum over $k\ne 0$) give $\sim\sinh^{-2}(\pi Rq)$ which
is exponentially suppressed for large $q$ and renders the integration
finite. Performing the integration in (\ref{two}) we find
\begin{equation}
\Pi_{55}^{AA'}(0)=-\frac{9}{32\pi^4R^2}g^2\delta^{AA'}C_2({\cal G})\zeta(3).
\label{gcHmass}\end{equation}
It is a manifestly finite result, as expected, since it is
the same result as the one we would have obtained 
in an $S^1$ compactification.

The $m$-dependent terms and therefore all terms involving $\partial_5$
derivatives vanish, which can be seen by writing Eq.~(\ref{three})
for $m\neq0$ as
\begin{equation}
\frac{{q^2+\frac{{l(m-l)}}{{R^2}}}}{ 
{(q^2+\frac{l^2}{R^2})(q^2+\frac{(m-l)^2}{{R^2}})}}=\frac{R}{m}
\left(
\frac{\frac{l}{R}}{{q^2+\frac{l^2}{R^2}}}-\frac{\frac{{l-m}}{R}}{ 
{q^2+\frac{(m-l)^2}{{R^2}}}}\right)
.\end{equation} 
Shifting the summation index $l\to l+m$ in the second term  cancels the
contribution from the first one.

Let us now compute $\Pi_{\mu\mu'}$. As in the case of $\Pi_{55}$ the
contributions of $\Pi^{(2,3)_{+}}$ and $\Pi^{(2,3)_{-}}$ effectively
add, yielding a total of
\begin{equation} \Pi_{\m\m'}^{AA'}(m^2)=
\frac{1}{2}g^2\delta^{AA'}C_2({\cal G}) 
\m^{4-d}\int \frac{d^{d }q}{{(2\pi)^{d}}}
\sum_{l=-\infty}^{+\infty} \frac{(1-d)\Bigl(
\frac{1}{d}(d-2)q^2+\frac{l^2}{R^2}\Bigr)+\frac{1}{2}
\frac{(5m^2-2ml)}{R^2}}{(q^2+\frac{l^2}{R^2})(q^2+\frac{(m-l)^2}{R^2})}.
\label{Pimubulk}\end{equation}
Performing the integral first we expect a quadratic divergence by
power counting. However, these divergences again cancel. This can be
seen in dimensional regularization by noticing that the coefficient of
the pole is proportional to $d-2$. Here the factor of $d-2$ is
necessary because the integral that it multiplies is not zero.  The
logarithmically divergent part however is now non zero for $m\ne 0$.
It fact it gives a contribution to the mass renormalization of the
heavy modes of the gauge bosons which is expected since the KK modes
being massive should not be protected by gauge invariance against such
divergences.  On the other hand for $m=0$ this would be a contribution
(together with (\ref{fig11})) to the mass of the massless ${\cal H}$
gauge bosons which should be forbidden by the gauge invariance of the
zero mode sector. Indeed, for $m=0$ (\ref{Pimubulk}) reduces to
\begin{equation} \Pi_{\m\m'}^{AA'}(0)=
\frac{1}{2}g^2\delta^{AA'}C_2({\cal G}) 
\m^{4-d}\int \frac{d^{d }q}{{(2\pi)^{d}}}\sum_{l=-\infty}^{+\infty}\Bigl(
(1-d)\frac{1}{{q^2+\frac{l^2}{R^2}}}-\frac{2}{d}(1-d)\frac{q^2}{ 
{(q^2+\frac{l^2}{R^2})^2}}\Bigr),\label{Pimubulk0}\end{equation}
and one can check that the whole expression is zero in any $d$.
 
Alternatively, we could have re-summed first and separated the five
from the four dimensional part from the beginning.  The five
dimensional part would then have a pole at $d=1$ (corresponding to a
cubic divergence in $d=5$).  Then, it is the factor of $1-d$ in
(\ref{Pimubulk}) that would protect against these divergences, while
the remaining integration would be found to be logarithmically
divergent for $m\ne 0$ and zero for $m=0$.

\subsubsection{\sc Brane effects from the gauge sector}
\label{braneeffects}

We will here gather brane effects from the brane sector results of
sections~\ref{thetadpole} and \ref{thegauge}.
We have to add the different contributions to $\widetilde\Pi_{55}$
according to Eq.~(\ref{SumPitilde}). The contributions are taken from
Eqs.~(\ref{tadpole}) and (\ref{gaugeandghost}) together with
Eqs.~(\ref{Ntilde25}) and (\ref{Ntilde35}):
\begin{equation}
\widetilde\Pi^{AA'}_{55}(m,m')=-\frac{d}{8}g^2\d^{AA'}
\bigl(2C_2({\cal H}_A)-C_2({\cal G})\bigr)(\eta^A+1)
\m^{4-d}\int \frac{d^{d }q}{{(2\pi)^{d}}}
\left(
\frac{1}{{q^2+m_-^2}}+
\frac{1}{{q^2+m_+^2}}
\right)\label{Pi55},
\end{equation}
where we have defined $m_-=\frac{1}{2R}(m-m')$ and $m_+=\frac{1}{
2R}(m+m')$. The last expression seems to indicate a quadratic
divergence. However, a closer look shows that this is not the case. To
see this, observe that $\widetilde\Pi^{AA'}_{55}(m,m')$ in
(\ref{Pi55}) is an even function of $m$ (and of $m'$) and therefore
the sum
\begin{equation} 
\sum_{m,m'} A^{a,m}_5\widetilde\Pi^{aa'}_{55}(m,m')A^{a,m'}_5=0,
\end{equation}  
since only gauge components with $\eta^A=+1$ contribute to brane
effects and therefore we are considering only negative parity scalar
fields $A^{a,m}_5$.  Thus, here we find a different reason for the
absence of divergences from the one we found in the bulk. There, the
poles canceled between the tadpole, the gauge and the ghost loop, as
it happens in a $d=4$ gauge theory; here they are simply zero.

Next let us examine the brane terms for the gauge bosons $M=\mu$.  We
find from Eqs.~(\ref{tadpole}) and (\ref{gaugeandghost}) together with
Eqs.~(\ref{Ntilde2mu}) and (\ref{Ntilde3mu})
\begin{multline}
\widetilde\Pi^{AA'}_{\mu,\mu'}(m,m')=-\frac{1}{16}g^2\d^{AA'}\delta_{\mu\mu'}
\bigl(2C_2({\cal H}_A)-C_2({\cal G})\bigr)(\eta^A+1)
\m^{4-d}\int \frac{d^{d }q}{{(2\pi)^{d}}}\\
\left(
\frac{\frac{4}{d}(d-2)(d-3)q^2+3\frac{m^2+m^{\prime 2}}{R^2}+
(d-2)\frac{(m+m')^2}{R^2}}{(q^2+m_-^2)(q^2+m_+^2)}
\right)\label{brg}.
\end{multline}
The first term is quadratically divergent but the pole cancels in
$d=2$ as expected. Notice that here the divergence is quadratic (since
there is no sum over $l$) so the poles should necessarily cancel in
$d=2$.  The brane mass term according to (\ref{braneexpansion}) is the
zero'th order of the above in an $m,m'$ expansion evaluated at
$m=m'=0$. It is simply
\begin{equation} 
\widetilde\Pi^{AA'}_{{\mu,\mu'}}(0,0)=0
\end{equation} 
as can be checked by doing the integral explicitly in dimensional
regularization. The rest of the terms in (\ref{brg}) correspond to a
logarithmically divergent piece and a finite piece. Recall that this
last computation corresponds to the vacuum polarization of the
unbroken ${\cal H}$ gauge bosons, so naively one would have guessed
that not only one should find no quadratic divergences but moreover
the logarithmically divergent and finite parts should also be
absent. What seems even more surprising is that the above amplitude is
non zero for say $m'=0$ and $m\ne 0$ which implies a mixing between
the zero mode and the heavy modes thus apparently breaking gauge
invariance.  We believe that the resolution to this puzzle is similar
to the one in the Standard Model where the mixing between the photon
and the $Z$ gauge boson computed in an $R_{\xi}$ gauge is non zero and
gauge non-invariant until vertex and box contributions are taken into
account.  We will not pursue this question any further, since
renormalization of these theories is not the main topic of this work.

\subsection{\sc The fermion sector}
\label{fermion2}

\begin{figure}[t]
\begin{center}
\quad 
\begin{picture}(120,60)(0,0)
\ArrowArcn(60,20)(20,0,180)
\ArrowArcn(60,20)(20,180,360)
\Photon(0,20)(40,20)23 
\Photon(80,20)(120,20)23 
\footnotesize
\Text(60,6)[c]{$r$}
\Text(60,33)[c]{$q$}
\Text(0,28)[l]{$m,A$}
\Text(120,28)[r]{${m^\prime},A'$}
\Text(41,0)[r]{$k,C$}
\Text(41,40)[r]{$l,B$}
\Text(79,0)[l]{$k',C'$}
\Text(79,40)[l]{$l',B'$}
\LongArrow(5,16)(20,16)
\Text(12,10)[c]{$p$}
\LongArrow(100,16)(115,16)
\Text(108,10)[c]{$p$}
\end{picture}
\end{center}
\caption{The diagram contributing from the fermion sector}
\label{fermiondiagram}
\end{figure}
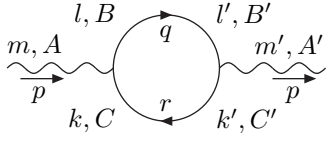

The diagram contributing from fermions living in the representation
${\bf R}$ is given in \fig{fermiondiagram} and evaluates to
\begin{align}
-(-ig)^2\delta_{m-l+k}\delta_{l'-k'-m'}
{\rm tr}\Bigl\{
&\frac{-i}{q_\r\g_\r+\frac{l}{R}\g_5}
	\frac{1}{2}\left(\delta_{l-l'}-\l_R\otimes (i\g^5)\d_{l+l'}\right)
	\g_{M'}T^{A'}_R\nonumber\\
&\frac{-i}{q_\s\g_\s+\frac{k'}{R}\g_5}
	\frac{1}{2}\left(\delta_{k'-k}-\l_R\otimes (i\g^5)\delta_{k'+k}\right)
\g_MT^A_R\Bigr\}.
\end{align}
Expanding the latter and using the identities (\ref{identitiesfirst})
$-$(\ref{identitieslast}), we obtain the momentum conserving terms
\begin{align}
\Pi_{MM'}^{(4)_+AA'}&=-\frac{g^2}{4}C_R\d^{AA'}
	\frac{1}{q^2+\frac{l^2}{R^2}}\frac{1}{q^2+\frac{(l-m)^2}{R^2}}
	{\rm tr}\left\{\g_R\g_{M'}\g_S\g_M\right\}q_Rr_S,\\
\Pi_{MM'}^{(4)_-AA'}&=-\frac{g^2}{4}C_R\eta^A\d^{AA'}
	\frac{1}{q^2+\frac{l^2}{R^2}}\frac{1}{q^2+\frac{(l-m)^2}{R^2}}
	{\rm tr}\left\{\g_R(i\g_5)\g_{M'}\g_S(i\g_5)\g_M\right\}q_Rr_S\a_S,
\end{align}
where $q_R=(q_\r,l/R)$, $r_S=(q_\sigma,k/R)$.
Using $\g_5\g_S=-\a_S\g_S\g_5$ together with Eq.~(\ref{SumPi}) gives
\begin{equation}
\Pi_{MM'}^{(4)AA'}=-\frac{g^2}{2}C_R\d^{AA'}
	\frac{1}{q^2+\frac{l^2}{R^2}}\frac{1}{q^2+\frac{(l-m)^2}{R^2}}
	{\rm tr}\left\{\g_R\g_{M'}\g_S\g_M\right\}q_Rr_S.
\label{fermionloopmc}
\end{equation}
The momentum violating terms are
\begin{align}
\widetilde\Pi_{MM'}^{(4)_+AA'}&=\frac{g^2}{4}{\rm tr}(\l_RT^{A'}_RT^A_R)
	\frac{1}{q^2+m_+^2}\frac{1}{q^2+m_-^2}
	{\rm tr}\left\{\g_R(i\g_5)\g_{M'}\g_S\g_M\right\}q_R^-q_S^+\a_S,\\
\widetilde\Pi_{MM'}^{(4)_-AA'}&=\frac{g^2}{4}{\rm tr}(T^{A'}_R\l_RT^A_R)
	\frac{1}{q^2+m_+^2}\frac{1}{q^2+m_-^2}
	{\rm tr}\left\{\g_R\g_{M'}\g_S(i\g_5)\g_M\right\}q_R^+q_S^-,
\label{fermionloopmv}
\end{align}
where here $q_R^+=(q_\r,m_+)$, $q_S^-=(q_\s,m_-)$.  These are
potential brane terms for any value of $\eta^A$.  However, one can
easily see that they are zero because of their gamma matrix structure:
\begin{equation} {\widetilde \Pi}_{55}^{(4)_{\pm}}(m,m')=0, \hskip 1cm 
{\widetilde \Pi}_{\m\m'}^{(4)_{\pm}}(m,m')=0.\end{equation}

The bulk contribution (\ref{fermionloopmc}) will be first evaluated
for $M=M^\prime=5$:
\begin{equation}
\Pi^{(4)AA'}_{55}(m^2)=
\frac{2^{\left[\frac{d}{2}\right]}}{2}g^2C_R{\d^{AA^\prime}}
{\m}^{4-d}\int \frac{d^{d }q}{{(2\pi)^{d}}}
\sum_{l=-\infty}^{+\infty}
\frac{{q^2+\frac{{l(m-l)}}{{R^2}}}}{{(q^2+\frac{l^2}{R^2})
(q^2+\frac{(m-l)^2}{R^2})}}.
\end{equation}
where $[x]$ is defined as usual as the integer part of $x$.  In
(\ref{two}) we saw that this integral is divergence free and that the
only non vanishing contribution comes from $m=0$.  Thus, the result
after summing and integrating, becomes
\begin{equation} 
\Pi^{(4)AA'}_{55}(m^2\ne 0)=0 \hskip 1cm {\rm and} \hskip 1cm 
\Pi^{(4)AA'}_{55}(0)=\frac{12}{32\pi^4R^2}g^2\d^{AA'}C_R\zeta(3)
.\label{fermmass}\end{equation}
We stress that here we did not have factors of $d-2$ ($d-1$) to
protect us against quadratic (cubic) divergences so it is fortunate
that this contribution is completely finite.

After a similar calculation, we obtain for the components along ${\cal M}_4$
\begin{equation} 
{\Pi}^{(4)AA'}_{\m\m'}(m^2)=\frac{2^{\left[\frac{d}{2}\right]}}
{2}g^2\d_{\m\m'}{\d^{AA^\prime}}C_R
{\m}^{4-d}\int \frac{d^{d }q}{{(2\pi)^{d}}}
\sum_{l=-\infty}^{+\infty}
\frac{{-\frac{1}{d}(d-2)q^2+\frac{{l(m-l)}}
{{R^2}}}}{{(q^2+\frac{l^2}{R^2})
(q^2+\frac{(m-l)^2}{R^2})}}.
\label{fermtot}\end{equation}
There is again no quadratic divergences since the corresponding pole
vanishes in dimensional regularization in $d=2$.  The logarithmic
divergences for $m\ne 0$ correspond to mass renormalization of the
heavy KK gauge bosons just as in the contribution (\ref{Pimubulk})
from the gauge sector.  For $m=0$ though these logarithmic divergences
are absent, since for $d=4$ the integral is proportional to
(\ref{Pimubulk0}) which was found to be zero, i.e.
\begin{equation}
\label{zeroferm}
{\Pi}^{(4)AA'}_{\m\m'}(0)=0\ .
\end{equation}

As a final consistency check, let us see how does the cancellation
of the pole work
if we perform the sum over $l$ first. As we have said earlier, 
cubic divergences which should somehow cancel will typically appear.
Indeed, we can rewrite (\ref{fermtot}) as
\begin{equation}
{\Pi}^{(4)AA'}_{\m\m'}(m^2)=\frac{2^{\left[\frac{d}{2}\right]}}
{2}g^2\d_{\m\m'}{\d^{AA^\prime}}C_R
{\m}^{4-d}\int \frac{d^{d }q}{{(2\pi)^{d}}}
\sum_{l=-\infty}^{+\infty}
\frac{{\frac{2}{d}(1-d)q^2+\bigl(q^2+\frac{{l(m-l)}}{
{R^2}}\bigr)}}{{(q^2+\frac{l^2}{R^2})
(q^2+\frac{(m-l)^2}{R^2})}}.
\end{equation} 
We have seen in Eq.~(\ref{three}) that the term in the bracket
corresponds to just a single finite contribution from $m=0$. The first
term is the one that has the cubic pole but it is multiplied by the
factor of $1-d$ so that the cubic divergence in $d=5$ is actually
absent.

\section{\sc The Hosotani mechanism}
\label{hosotani}

In this section we will discuss on the possibility that one of the
scalars $A_5^{{\hat a},0}$ radiatively acquires a vacuum expectation
value (VEV) and breaks the gauge group ${\cal H}$ on the brane to a
subgroup.  This would be relevant in model building for example when
$SU(2)_W\times U(1)_Y$ is a subgroup of ${\cal H}$, the scalars
$A_5^{{\hat a},0}$ are interpreted as Higgs fields and the VEV
\begin{equation} \o^{\hat a}\equiv \frac{1}{2}\langle A_5^{{\hat a},0}
\rangle R \end{equation}
breaks ${\cal H}$ down to ${U(1)}_{Q}$.  This mechanism is called the
Hosotani mechanism \cite{Hosotani:1983xw, Hosotani:1989bm}. It is not
easy to carry out a discussion as general as the one we had up to this
point so we will make a few simplifying assumptions. First, we will
assume that there is only one type of fermions, transforming either in
the adjoint or in the fundamental representation of ${\cal G}$. We
allow though for multiple flavors of fermions and we will call the
number of different flavors by $N_f$. Second, we assume that only one
of the Higgs fields $A_5^{{\hat a},0}$ takes a VEV, we will call this
field $h$ and write $\o=v R/2$, where $v\equiv \langle h\rangle$.

The first step is to look at the squared mass at the origin of the
Higgs field. Adding the contribution (\ref{fermmass}) multiplied by
the number of fermion flavors to the result (\ref{gcHmass}) from the
gauge sector and noticing that in Euclidean space we are computing the
negative mass squared, we obtain the result
\begin{equation} {m_{h}^2}=
\frac{3}{32\pi^4R^2}g^2\zeta(3)
\bigl(3C_2({\cal G})-4C_RN_f\bigr).\label{Higgsmass}\end{equation} 
We see that for models satisfying (we use $C_2({\cal G})=C_{\cal G}$)
\begin{equation} \frac{C_{\cal G}}{C_R} < \frac{4}{3}N_f\label{cond0}\end{equation}
this is negative and therefore $\o$ could indeed break ${\cal H}$.  Of
course, even when $m_h^2>0$ the true vacuum can be at some nonzero
$\o$.  To be more precise, we have to look at the full effective
potential which can be expressed as~\cite{Delgado:1998qr}
\begin{equation} 
V=\frac{1}{128\pi^6R^4}Tr\Bigl(V(r_F)-V(r_B)\Bigr),
\end{equation} 
with
\begin{equation} 
V(r)=3(Li_5(r)+Li_5(r^*)), \hskip 1cm r_{F,B}=e^{2\pi iq_{F,B}(\o)}
\end{equation}
and where $q_F$ and $q_B$ are the shifts in the fermion and boson KK
masses according to 
\begin{equation} 
m_{F}^n=\frac{n+q_F(\o)}{R},\hskip 1cm 
m_{B}^n=\frac{n+q_B(\o)}{R}, \hskip .5cm n=0,\pm 1, \pm 2, \cdots.
\end{equation}
The question then is whether this potential has a minimum for some
non-zero value of $\o$, which would then trigger the breaking of
${\cal H}$, and if this minimum is a global minimum. It is hard to
answer this question in full generality but one can make some progress
in special cases.  Two simple examples have been provided in
Ref.~\cite{Kubo:2001zc}.  For $N_f$ fermions in the fundamental
representation of $SU(2)$ it was found that there is a global minimum
at $\o=\frac{1}{2}$~\footnote{Note that the parameter $\a$ in
Ref.~\cite{Kubo:2001zc} is related to our $\o$ as $\a=2\o$.} which
becomes degenerate with the one at $\o=0$ when $N_f\rightarrow 0$.
Thus, there is a regime\footnote{According to Eq.~(\ref{cond0}),
$\o=0$ is a maximum for $N_f\geq 3$.} $0<N_f\leq 3$ where the minimum
at $\o=0$ does not correspond to the true vacuum. However it was also
pointed out that the $U(1)$ symmetry generated by $\s^3$ which is left
over after the orbifolding remains unbroken. This is because the
Wilson-line associated to the vacuum $\o=\frac{1}{2}$ becomes $-\One$
which commutes with $\s^3$. In the other example, ${\mathcal G}=SU(3)$
with fermions in the fundamental, it was shown that there again exists
a minimum at $\o=\frac{1}{2}$, which already becomes the true vacuum
when $N_f>\frac{3}{2}$. This is a less stringent bound than
Eq.~(\ref{cond0}), $N_f > \frac{9}{2}$, the critical value at which
the minimum at $\o=0$ turns into a maximum. By computing the Wilson
line associated to $\o=\frac{1}{2}$ it was finally shown that this
breaks only the $SU(2)$ subgroup of $\mathcal H$ down to $U(1)$. As in
the case of $SU(2)$ the rank remains preserved.

If the fermions transform in the adjoint representation of $\mathcal
G$ there is a slight simplification because $q_F=q_B$. The $\o$
dependence becomes the same for fermionic and bosonic contributions
and the symmetry breaking is determined by the global factor $3-4N_f$
in front of the potential. For $SU(2)$ one finds two degenerate minima
which lie at $\o=0, \frac{1}{2}$ for $N_f<\frac{3}{4}$ and at
$\o=\frac{1}{4}, \frac{3}{4}$ for $N_f>\frac{3}{4}$. For $SU(3)$ one
has only a minimum at $\o=0$ for $N_f<\frac{3}{4}$ and two degenerate
minima at $\o\simeq 0.29,0.71$ for $N_f>\frac{3}{4}$.

Our next observation is the following statement: The Hosotani
mechanism does not reduce the rank of ${\cal H}$ if the symmetry
breaking global minimum is at $\o=\frac{1}{2}$.  To show this
statement, we can compute the Wilson line due to the VEV $\frac{1}{R}$
of a scalar along the $T^A$ direction:
\begin{equation} \langle W\rangle =e^{i\pi T^{A}}.\end{equation} 
It is straightforward to show that $\exp({i\pi T^{A}})$ is a diagonal
matrix.  Thus, we always have that
\begin{equation}  [\langle W\rangle ,H_i] =0,\end{equation}
where $H_i$ are the generators corresponding to the Cartan subalgebra
of ${\cal G}$, i.e. that the Wilson loop commutes with at least those
generators and therefore it leaves at least a $U(1)_1\times \cdots
\times U(1)_{rank({\cal H})}$ unbroken.  On the other hand if $\o\neq
\frac{1}{2}$ then one can reduce the rank of ${\cal H}$ by Hosotani
breaking. An example of this possibility was provided above where
$N_f>\frac{3}{4}$ fermions in the adjoint representation of $SU(N)$
triggered a Hosotani breaking with $\o\neq \frac{1}{2}$.

One should however be careful with the interpretation of this effect.
Let us look at the example of the orbifold breaking $SU(2)\stackrel
{orb}{\longrightarrow} U(1)$, where the subsequent Hosotani mechanism
according to the above argument apparently leaves the gauge group
unbroken when $\o=1/2$: $U(1)\stackrel {Hos}{\longrightarrow} U(1)$.
Strictly speaking, this is not correct.  By looking at the symmetry
breaking pattern in detail it can be seen that the gauge boson
$A^{3,0}_{\m}$ of ${\cal H}=U(1)$ becomes massive by the Hosotani
vacuum expectation value and therefore ${\cal H}$ breaks to nothing on
the brane for generic values of $\o$. However, for special values of
$\o$, such as the value $\o=1/2$ corresponding to fermions
transforming in the fundamental representation of $SU(2)$, additional
gauge bosons can become massless, as can be seen from the mass matrix
of the massive KK gauge bosons which has eigenvalues $n^2, (n+2\o)^2,
(n-2\o)^2$. Clearly, since $n>0$, the gauge boson associated to the
$n=1$ level (some linear combination of $A^{1,1}_{\m}$ and
$A^{3,1}_{\m}$ in this case) becomes massless, so the correct
statement is $U(1)\stackrel {Hos}{\longrightarrow} U(1)'$.  In fact,
in general, for special values of $\o$ there will be an enhanced gauge
symmetry at the orbifold fixed points~\footnote{Gauge symmetry
enhancement at orbifold fixed points is a known effect in string
theory.}.  The interesting feature here is that the special value of
$\o$ that results in gauge symmetry enhancement is not arbitrary,
instead it is obtained by minimizing the fixed (once a bulk fermion
representation is chosen) effective potential. Recall that $\o\sim R
v$ and that having decoupled gravity, $R$ is a parameter assumed to be
fixed to some reasonable value. Then, $v$ is essentially the (only)
classical modulus of the theory. In supersymmetric theories $v$ might
remain a modulus even at the quantum level but in non-supersymmetric
theories such as the ones we analyze here, it can be fixed through a
non-trivial one loop potential as we have seen above.
       
It would be interesting to see if, by turning on gravity, $R$ could be
stabilized in the same manner. Of course, in such a case, another
important issue would be if $R$, being the extra component of the
metric (i.e. $g_{55}$), is protected by gauge invariance against
quadratic divergences like $A_5^{\hat a}$, gauge invariance now being
general coordinate invariance.

\section{\sc Conclusions and outlook}
\label{conclusion}

In this paper we have analyzed the one-loop bulk and brane induced
 radiative effects in a higher dimensional gauge theory compactified
 on an orbifold that breaks an arbitrary gauge group $\mathcal{G}$
 into its subgroup $\mathcal{H}$.  We have restricted our explicit
 analysis to a five-dimensional theory compactified on
 $\mathcal{M}_4\times S^1/\mathbb{Z}_2$.  The gauge group at the fixed
 point branes is $\mathcal{H}$ and the higher dimensional components
 of the gauge bosons along $\mathcal{G}/\mathcal{H}$ are even scalar
 fields whose (massless) zero modes can become tachyonic by radiative
 corrections in the bulk and trigger spontaneous Hosotani breaking of
 $\mathcal{H}$, i.e. they play the role of Higgs fields with respect
 to the gauge group $\mathcal{H}$.

Mass renormalization of Higgs fields in the bulk is expected to be
protected from quadratic divergences by the higher dimensional gauge
invariance $\mathcal{G}$ of the theory. However, since the branes
localized at the fixed points are four dimensional space-times, mass
renormalization of the Higgs fields on the branes is not a priori
protected from quadratic divergences by the higher dimensional theory.

We have computed the one-loop mass renormalization of the Higgs (and
gauge) fields in the bulk and on the branes and found no quadratic
divergences at all for any of them. While this effect in the bulk is
justified from the higher dimensional gauge invariance its
interpretation on the brane for the Higgs fields is less clear,
although we believe it might be related to the higher-dimensional
Lorentz and gauge symmetries.  In fact, we have used a five
dimensional covariant gauge (Feynman-'t Hooft gauge) and dimensional
regularization that are both consistent with the higher dimensional
gauge invariance.

In particular, for the Higgs fields all brane 
effects vanish while bulk effects are finite and can trigger, depending
on the fermionic content of the theory, spontaneous Hosotani breaking of
the brane gauge group $\mathcal{H}$. For the gauge fields we find only
logarithmic divergences consistent with the mass renormalization of heavy
KK modes.

Our results are also consistent with the Higgs fields acquiring a VEV
and thus spontaneously break the gauge symmetry $\mathcal{H}$ on the
branes with one-loop insensitivity to the ultraviolet cutoff of the
higher dimensional theory. This insensitivity seems to be a remnant of
the properties of the higher dimensional theory where the brane is
embedded in.  Our results prove that the higher dimensional gauge
theory provides a one-loop solution to the hierarchy problem, i.e. it
replaces the cutoff by the scale $1/R$ above which the effective
theory becomes higher dimensional, modulo the stabilization of the
compactification radius that should involve the gravitational sector
of the theory. In any case, a quadratic divergence on the brane would
have recreated the hierarchy problem.

Of course our framework cannot be considered as a full solution to the
absence of quadratic divergences until they are proved to vanish at
any order of perturbation theory, or the symmetry protecting them is
clearly identified. In fact, a naive analysis of two-loop diagrams
prove that there should be non-vanishing Higgs mass renormalization
brane effects from two-loop diagrams. However, since there are
one-loop wave function renormalization effects localized on the
branes, they should induce on finite one-loop mass diagrams
non-vanishing two-loop effects localized on the branes. This means
that two-loop mass renormalization effects on the branes are
mandatory. Only a genuine two-loop calculation can disentangle a
two-loop effect induced by the wave function renormalization localized
on the brane (and thus absorbable by renormalization group running as
happens in supersymmetric theories) from a genuine Higgs mass
counter-term localized on it. A two-loop calculation is beyond the
scope of the present paper but will be the subject of future
investigation by the present authors.

\section*{\sc Acknowledgments} 
The work of GG was supported by the DAAD.

\appendix

\section{\sc Orbifold gauge breaking patterns}
\label{Appendix}

We will not try to derive general rules for the allowed gauge breaking
patterns and the associated consistent fermion
representations~\footnote{A treatment of parity assignments to pure
gauge-theories and a list of all possible breaking patterns can be
found in Ref.~\cite{Slansky:1981yr, Hebecker:2001jb}.}.  Most of the
general features of orbifold actions on gauge fields and fermions have
appeared in one way or another in the early string theory literature.
Instead, in this appendix, we work out a few examples demonstrating
the simplicity but also the restrictiveness of the $\mathbb{Z}_2$
orbifold action on the fermion representations.  We recall that the
gauge group in the bulk is ${\cal G}$, which breaks by the orbifold
action represented by
\begin{equation} \L = \left(\begin{array}{cc} {\bf 1}_{d_{\cal H}} & 0 \\ 
0 & -{\bf 1}_{d_{\cal K}} \end{array}\right), \hskip 1cm
{\rm such~ that} \hskip .5cm f^{ABC}=\L^{AA'}\L^{BB'}\L^{CC'}f^{A'B'C'},
\label{automorphism1A}\end{equation}
as 
\begin{equation} {\cal G}\rightarrow {\cal H},\end{equation}
where ${\cal H}={\cal H}_1\otimes {\cal H}_2\otimes\cdots$
and the broken generators parametrize the coset 
${\cal K}={\cal G}/{\cal H}$. 
The fermion representation of the bulk then breaks up 
according to this, as
\begin{equation} {\bf R}={\bf r}_1\oplus {\bf r}_2\oplus \cdots,\end{equation}
where the ${\bf r}_i$ have to be determined by 
the orbifold action on the fermions 
\begin{equation} \Psi_R(x^{\m},-x_5)=\l_R \otimes (i\g^5)\Psi_R(x^{\m},x^5) 
\label{pfermionapp}\end{equation} 
and from the requirement that
the coupling $igA_{M}^A{\overline \Psi_R}\g^{M}T^A{\Psi_R}$
is $\mathbb{Z}_2$ invariant. The resulting constraint from the latter is 
\begin{equation} \l_R T_R^{A} \l_R=\eta^A T_R^A, 
\label{automorphism2A} \end{equation} 
which can be simplified to
\begin{equation} [\l_R,T_R^a]=0\hskip 1cm \{\l_R,T_R^{\hat a}\}=0 \label{auto2}. \end{equation} 
To put it in simple words, for a given representation ${\bf R}$,
$\l_R$ has to be chosen in such a way that it commutes with the
unbroken generators $T_R^a$ and anti-commutes with the broken
generators $T_R^{\hat a}$.

The first class of models of interest is when $\L$ is such that
$rank({\cal G})=rank({\cal H})$ (inner automorphism).  In particular
this means that none of the $T_R^{\hat a}$ is diagonal and therefore
$\l_R$ can always be chosen to be diagonal of the form
\begin{equation} 
\l_R = \left(\begin{array}{cc} {\bf 1}_{d_1} & 0 \\ 
0 & -{\bf 1}_{d_2} \end{array}\right)
,\end{equation}
where $d_1$ and $d_2$ are model dependent numbers.  This is not the
case for the second interesting class of models, the one with $\L$
chosen such that the rank of ${\cal G}$ is reduced (outer
automorphism). Reduced rank in particular implies that some of the
$T_R^{\hat a}$ are diagonal and therefore (for those diagonal
$T_R^{\hat a}$) the second equality of (\ref{auto2}) can never be
satisfied if $\l_R$ is diagonal.  Thus, for the case of outer
automorphism we have to find a non-diagonal (and unitary) $\l_R$ that
solves (\ref{auto2}).  The most interesting case will be
\begin{equation} 
\Lambda^{AB} T^B=-\left( T^A \right)^T, \label{conjugate}
\end{equation} 
which e.g.~breaks $SU(N)\rightarrow SO(N)$.  From
Eq.~(\ref{automorphism2A}) it then follows that possible
representations must be real:
\begin{equation}
-\left( T_R^A \right)^T=\lambda_R T_R^A \lambda_R\label{real}
\end{equation}
(recall that $\l=\l^\dagger=\l^{-1}$).  For nonreal representations
${\mathcal R}$ one can always choose ${\bf R}={\mathcal R}\oplus
\overline{{\mathcal R}}$ with generators
\begin{equation}
T^A_{R}=\left(\begin{array}{cc} T^A_{\mathcal R} & 0 \\ 
0 & -\left( T_{\mathcal R}^A \right)^T\end{array}\right).
\end{equation}
Comparing with Eq.~(\ref{real}) this fixes $\lambda_{R}$ to take the
block form
\begin{equation}
\lambda_{R}=\left(\begin{array}{cc} 0 & 1_{\mathcal R} \\ 
1_{\mathcal R} & 0 \end{array}\right).\label{nondiagonallambda}
\end{equation}
Thus $\lambda_{\mathcal R}$ has $d_{\mathcal R}$ eigenstates $(\hat
e_i,\hat e_i)$ with positive parity and $d_{\mathcal R}$ eigenstates
$(\hat e_i,-\hat e_i)$ with negative parity, where with $\hat e_i$ we
denote the usual unit vectors.  The zero mode spectrum resulting from
this action will always be vector-like and therefore anomaly free.
Let us now present a few examples for both the rank preserving and the
rank breaking orbifold actions.

The first example in the inner automorphism class is an $SU(2)$ gauge
group in the bulk with a pair of Dirac fermions transforming as a
doublet under the fundamental representation of $SU(2)$.  First, we
have to choose the action $\L$ on the gauge fields. We have
essentially three choices. The first is to take $\L = diag(+1,+1,+1)$.
This choice corresponds to an unbroken $SU(2)$ on the fixed planes and
therefore it is not an interesting choice from our point of view.  The
second possibility is to take $\L = diag(-1,-1,-1)$ which corresponds
to a completely broken gauge group on the fixed planes.  However, we
have seen in section 2 that this choice is not compatible with the
automorphism constraint on the Lie algebra so this possibility cannot
be realized.  The third choice is to choose $\L=diag(-1,-1,+1)$, which
breaks $SU(2)$ down to $U(1)$ on the brane.  The gauge boson of
positive parity corresponding to the unbroken $U(1)$ is $A_{\m}^3$,
whereas the broken coset is spanned by the negative parity
$A_{\m}^{1,2}$. Similarly, the zero modes of the positive parity
$A_5^{1,2}$ are seen as massless scalars in the four dimensional
theory but the negative parity $A_5^{3}$ does not have a zero mode.
This is an interesting possibility, so let us look at the action on
the fermions in the fundamental representation ${\bf R}={\bf 2}$ of
$SU(2)$.  One can easily check that $\l_{\bf 2}=diag(+1,-1)$
($d_1=d_2=1$) commutes with $T_R^3=\frac{1}{2}\s^3$ and anti-commutes
with $T_R^{\hat 1}=\frac{1}{2}\s^1$ and $T_R^{\hat
2}=\frac{1}{2}\s^2$.  The surviving fermions on the brane are then
(say) left handed Weyl fermions with $U(1)$ charge +1 and right handed
fermions with $U(1)$ charge $-1$. Thus, the orbifold action resulted
in a broken gauge group and chiral fermions on the brane.  The theory
on the brane is anomaly free as can be readily checked.

As a second example, let us look at the breaking pattern
$SU(3)\rightarrow SU(2)\otimes U(1)$.  To achieve this breaking
pattern, we take $\L=diag(+1,+1,+1,-1,-1,-1,-1,+1)$, so that the
generators corresponding to ${\cal H}=SU(2)\otimes U(1)$ have positive
parity and the generators corresponding to ${\cal
K}=SU(3)/(SU(2)\otimes U(1))$ have negative parity. We also choose the
fermions to transform in the fundamental representation of $SU(3)$,
i.e. ${\bf R}={\bf 3}$.  Then, $\l_{\bf 3}=diag(+1,+1,-1)$ ($d_1=2$,
$d_2=1$) commutes with the generators of ${\cal H}$ and anti-commutes
with the generators of ${\cal K}$, so on the brane we will have two
massless Weyl fermions transforming as a doublet under $SU(2)$ and a
singlet massless Weyl fermion of the opposite chirality.  Under
$SU(3)\supset SU(2)\otimes U(1)$ we have
\begin{equation} {\bf 3}={\bf 2}_{-1}\oplus{\bf 1}_{+2}
.\end{equation}
%
%
This model as it stands has a cubic $U(1)$ and an $SU(2)$ anomaly.
One has, in principle, to make modifications that render it 
anomaly free~\footnote{An example can be found e.g. in 
Ref.~\cite{Antoniadis:2001cv}.}.  

The third and last example in the inner automorphism class is an
$SU(5)$ in the bulk with fermions transforming under some
representation of the gauge group. We will not go through all the
possibilities, instead we look at the breaking pattern
$SU(5)\rightarrow SU(3)\otimes SU(2)\otimes U(1)$.  Following the
previous examples, we take $\L$ to have +1's along the diagonal
corresponding to the 12 generators of $SU(3)\otimes SU(2)\otimes U(1)$
and $-1$'s along the rest of the diagonal elements.  Let us assume
that the representation is ${\bf R}={\bf \overline 5}$.  Next we have
to look for a 5 by 5 matrix $\l_R$ that commutes with all the
generators of ${\cal H}$ and anti-commutes with the generators of
${\cal K}$. These constraints then fix $\l_{\bf 5}
=diag(+1,+1,+1,-1,-1)$ ($d_1=3$, $d_2=2$), since for $SU(5)\supset
SU(3)\otimes SU(2)\otimes U(1)$
\begin{equation} {\bf \overline 5}=
({\bf \overline 3},{\bf 1})_{+1/3}\oplus ({\bf 1},{\bf 2})_{-1/2}
.\end{equation}
The fermionic zero mode sector consists then of an $SU(3)$ triplet of
Weyl fermions of say right handed chirality, and an $SU(2)$ doublet of
Weyl fermions of left handed chirality.  It is also possible to carry
out the same exercise for ${\bf R}={\bf 10}$. Here it is more
convenient to express $\l_{\bf 10}$ in a tensor rather than a matrix
form since the ten is the antisymmetrized tensor product of the
fundamental with itself. We find
\begin{equation} {\l_{\bf 10}}^{il}_{~mn} =
\frac{1}{2}({\l_{\bf 5}}^i_{~m}{\l_{\bf 5}}^l_{~n}-
{\l_{\bf 5}}^i_{~n}{\l_{\bf 5}}^l_{~m}),\end{equation} 
which implies that in
\begin{equation} {\bf 10}=({\bf 3},{\bf 2})_{1/6}\oplus 
({\bf {\overline3}},{\bf 1})_{-2/3}\oplus ({\bf 1},{\bf 1})_{1}\end{equation}
the zero mode spectrum consists of a $({\bf 3},{\bf 2})$ of left
handed chirality and a $({\bf {\overline3}},{\bf 1})\oplus ({\bf
1},{\bf 1})$ of right handed chirality. For a model that has Dirac
fermions in the ${\bf \overline 5}\oplus {\bf 10}$ in the bulk, the
zero mode spectrum is clearly anomaly free, since the spectrum is that
of the Standard Model.

Before turning to examples of outer automorphisms, we would like to
make the connection between the general formalism reviewed
in~\cite{Hebecker:2001jb} and formulas (\ref{automorphism1A}) and
(\ref{automorphism2A}) again through simple examples. We recall that
the action of the orbifold group on the fields, when the action is an
inner automorphism, is via group elements such that
\begin{equation} g=e^{-2\pi i V\cdot H},\label{groupelem}\end{equation}
where $H=\{H_i\}, i=1,\cdots ,rank({\cal G})$  
are the generators of the Cartan subalgebra of ${\cal G}$  and
$V$ is the twist vector specifying the orbifold.
For such group elements it is always true that
\begin{equation} gH_ig^{-1}=H_i\hskip 1cm {\rm and}\hskip 1cm 
gE_{\a}g^{-1}=e^{-2i\pi \a\cdot V}E_{\a},\label{inner}\end{equation}
where $\a=\{\a_i\}$ are the roots of ${\cal G}$ and $E_{\a}$
the corresponding ladder generators in the Cartan-Weyl basis, satisfying
\begin{equation} [H_i,E_{\a}]=\a_iE_{\a}.\end{equation}
Let us first look at the $SU(2)\rightarrow U(1)$ example.  Taking the
Cartan generator of $SU(2)$ in the adjoint representation, i.e.
$H_1=T_A^3$ and requiring that $g=\L=diag(+1,+1,-1)$ in
(\ref{groupelem}), fixes $V=-1/2$. Then, $\l_{\bf 2}$ is given again
by (\ref{groupelem}) with $H_1=T_{\bf 2}^3$.  A simple calculation
yields $\l_{\bf 2}=diag(+1,-1)$ (up to a sign) as we had found
earlier. The exponent in the second of Eq.~(\ref{inner}) is non-zero
for all non-zero roots of $SU(2)$ ($\a=\pm 1$, so $\exp{(-2i\pi
\a\cdot V)}=-1$) and therefore the only unbroken generator is $H_1$.
A similar calculation for the $SU(3)\rightarrow SU(2)\otimes U(1)$
example gives $V=(0,\sqrt{3})$ and therefore $\l_{\bf
3}=diag(+1,+1,-1)$ (up to a sign).  In this case, however, from
(\ref{inner}) we can see that in addition to $H_1$ and $H_2$ there are
two more unbroken generators, namely $E_{\pm 1}$, corresponding to
$\a_{\pm1}=(\pm 1,0)$, since for those it is $\a_{\pm 1}\cdot V=0$.
For the rest of the generators $E_{\pm 2}$ and $E_{\pm 3}$ we find
$\exp{(-2i\pi \a\cdot V)}=-1$ as expected.

Finally, we will present two examples with rank breaking actions.  The
first example is the simplest possible one, namely a $U(1)$ in the
bulk breaking down to nothing on the branes.  The choice that performs
this breaking is $\L=-1$ and is a simple realization of
Eq.~(\ref{conjugate}). Charged fermions are necessarily accompanied by
oppositely charged partners.  Thus, the fermionic zero mode spectrum
is a vector-like pair of Weyl fermions.

The second simplest example with rank breaking action is
$SU(3)\rightarrow SO(3)$.  Solving Eq.~(\ref{conjugate}) to obtain
$\L$, one has to give positive parities to antisymmetric and negative
ones to symmetric generators. This results in the choice
\begin{equation}
\L=diag
(-1,+1,-1,-1,+1,-1,+1,-1).
\end{equation}
The parities for fermions are given by Eq.~(\ref{nondiagonallambda}).
Since the positive generators $T^2,\ T^5,\ T^7$ form the fundamental
representation of $SO(3)$, we find that matter in ${\bf 3}\oplus{\bf
\bar 3}$ of SU(3) will transform in ${\bf 3}\oplus{\bf 3}$ of SO(3).
We checked also that after diagonalizing $\lambda_{{\bf 3}\oplus{\bf
\bar 3}}$ the positive and negative parity eigenstates transform in
separate ${\bf 3}$ representations of $SO(3)$.  The fermionic zero
mode spectrum is therefore an $SO(3)$ triplet of Weyl fermions plus
their vector-like partners.  Finally we mention that other choices of
outer automorphisms not obtained from Eq.~(\ref{conjugate}) are
related to this one by an inner automorphism.

\section{\sc The unitary gauge}
\label{AppendixB}

In this appendix we will study the problem of gauge fixing and the
physical (unitary) gauge in the class of 5D models compactified on the
orbifold $S^1/\mathbb{Z}_2$ considered in this paper, where the gauge
group $\mathcal{G}$ is broken by the orbifold action into its subgroup
$\mathcal{H}$. As we have seen in section~\ref{hosotani} in the
presence of non-vanishing background values for the scalars
$A_5^{A,0}$ in the adjoint representation of $\mathcal{G}$ the
subgroup $\mathcal{H}$ can be further broken and the mass pattern
induced by the orbifold breaking will be modified. We will first
consider, for simplicity the case of zero VEV for $A_5^{A,0}$.  The
Hosotani breaking case will be subsequently studied.

We have seen that the choice of the gauge 
\begin{equation}
\label{gauge0}
G^A=\frac{1}{\sqrt{\xi}}\ \partial^M A_M^A
\end{equation}
does not lead, for any value of the parameter $\xi$, to the unitary
gauge.  In the absence of VEV for the fields $A_5^{A,0}$ this can be
achieved for the gauge fixing condition
\begin{equation}
\label{gauge1}
G^A=\frac{1}{\sqrt{\xi}}\ \left(\partial^\mu A_\mu^A + \xi \partial^5
A_5^A\right),
\end{equation}
consistent with the orbifold action.
Using now the gauge-fixing condition (\ref{gauge1}) and the
infinitesimal transformation of the field $A_M$ under the gauge
transformation $\alpha(x^\mu,x^5)$, $$A_M\to A_M+{\displaystyle
\frac{1}{g}} D_M \alpha,$$ standard techniques yield the Faddeev-Popov
Lagrangian 
\begin{equation}
\label{FP}
\mathcal{L}_{FP}=-\ \Tr\ \bar{c}\left(\partial^\mu D_\mu+\xi \partial^5
D_5\right)c.
\end{equation} 
The propagators can be worked out as in (\ref{gbprop}) and
(\ref{ghostprop}) and yield
\begin{equation}
\label{gbpropu}
G^{(A_\mu)}(p,p_5)=-i\frac{\delta^{BC}}{p^2-p_5^2}\left(g_{\mu\nu}-
\frac{(1-\xi)p_\mu p_\nu}{p^2-\xi p_5^2}\right),
\end{equation}
\begin{equation}
\label{g5propu}
G^{(A_5)}(p,p_5)=-i\frac{\delta^{BC}}{p^2-\xi p_5^2},
\end{equation}
\begin{equation}
\label{ghostpropu}
G^{(c)}(p,p_5)=-i\,\frac{\delta^{BC}}{p^2-\xi p_5^2}.
\end{equation}
After mode decomposition $p_5=n/R$ and so for $n\neq 0$ one reaches
the unitary gauge in the limit $\xi\to\infty$. In this limit the
massive modes of $A_5^A$ and $c^A$ decouple while the massive modes of
$A_\mu^A$ only propagate their physical degrees of freedom. This
corresponds to the gauge fixing condition $\partial^5 A_5^A=0$, that
does not fix the gauge in the zero-mode sector.  For the zero modes
left out by the orbifold breaking the gauge symmetry is unbroken and,
in the absence of Hosotani breaking, one cannot define a unitary
gauge, as it is obvious from Eqs.~(\ref{gbpropu}) to
(\ref{ghostpropu}).

Turning now VEVs for fields $A_5^{A,0}$ some of the massless gauge
bosons in $\mathcal{H}$ acquire a mass and the definition of the
unitary gauge can be enlarged to also take into account this
effect. The analysis can be readily done in full generality as
follows.  Let us consider that some fields $A_5^{A,0}$ will acquire a
VEV by quantum corrections. Their tree level potential is flat, since
they are part of the gauge bosons in 5D, and we can write the general
decomposition for them into a classical part and quantum fluctuations
as
\begin{equation}
\label{descomp}
A_5^{A,0}(x^\mu)=v^A+\chi^A(x^\mu).
\end{equation}
If we restrict ourselves to $x^\mu$-dependent gauge transformations
$\alpha= \alpha(x^\mu)$ we can move away from $v^A$ by means of gauge
transformations $\delta A_5^{A,0}={f^A}_{BC}A_5^{B,0} \alpha^C$ which
shows that field fluctuations along ${f^A}_{CB}v^C$ correspond to
Goldstone bosons for the zero mode sector. In fact, if we define the
mass matrix
\begin{equation}
\label{masa}
\mathcal{M}^A_B=g {f^A}_{CB}v^C,
\end{equation}
the zero modes of the gauge bosons acquire the squared mass matrix
\begin{equation}
\label{masa2}
\left(M^2_A\right)^{AB}=\left(\mathcal{M}\mathcal{M}^T \right)^{AB}.
\end{equation}
In order to incorporate the Hosotani
breaking into the 5D formalism we can modify the gauge fixing
condition (\ref{gauge1}) and define the $R_\xi$ gauge
\begin{equation}
\label{gauge2}
G^A=\frac{1}{\sqrt{\xi}}\ \left[\partial^\mu A_\mu^A + \xi \left(\partial^5
A_5^A+\mathcal{M}^A_B \chi^B\right)\right].
\end{equation}
The Goldstone bosons (zero modes) acquire similarly a mass
\begin{equation}
\label{masa3}
\left(M^2_G\right)_{AB}=\xi\left(\mathcal{M}^T\mathcal{M} \right)_{AB}.
\end{equation}
Of course, not only the zero modes will acquire a symmetry breaking
mass but also all massive modes will get the mass
\begin{equation}
\label{masatotal}
\left(\mathcal{M}^{n}\right)^A_B=m_n\delta^A_B+ \mathcal{M}^{A}_B,
\end{equation}
where $m_n=\pm n/R$ is the compactification mass and
$\mathcal{M}^{A}_B$ is the symmetry breaking mass given in
(\ref{masa}). Then propagators (\ref{gbpropu}), (\ref{g5propu}) and
(\ref{ghostpropu}) become
\begin{equation}
\label{gbpropuf}
G^{(A_\mu)}(p,n)=\left[\frac{-i}{p^2-\mathcal{M}^n
(\mathcal{M}^n)^T}\left(g_{\mu\nu}- \frac{(1-\xi)p_\mu
p_\nu}{p^2-\xi\mathcal{M}^n (\mathcal{M}^n)^T}\right)\right]^{AB},
\end{equation}
\begin{equation}
\label{g5propuf}
G^{(A_5)}(p,n)=-i\left[\frac{1}{p^2-
\xi(\mathcal{M}^n)^T\mathcal{M}^n}\right]_{AB},
\end{equation}
\begin{equation}
\label{ghostpropuf}
G^{(c)}(p,n)=-i\,\left[\frac{1}{p^2-\xi\mathcal{M}^n
(\mathcal{M}^n)^T }\right]^{AB}.
\end{equation}
The matrix character of the propagators implies that the matrix
$\mathcal{M}^n$ should be invertible, i.e. it satisfies the condition
\begin{equation}
\label{inversa}
\det\left[ \mathcal{M}^n\right]\neq 0,
\end{equation}
which is the necessary condition for gauging away the corresponding
Goldstone boson.

\bibliographystyle{h-elsevier}

\bibliography{/users/mariano/Nikos/refs}

\end{document}